\documentclass[%
 reprint,
prb,
floatfix,
]{revtex4-2}

\usepackage{blindtext}
\usepackage{graphicx}
\usepackage{dcolumn}
\usepackage{bm}
\usepackage{rotating}
\usepackage{amsmath}
\usepackage{booktabs}
\usepackage{tabularx}
\usepackage{appendix}
\usepackage{footmisc}
\usepackage{float}
\usepackage{placeins}
\usepackage{afterpage}
\begin{document}

\preprint{APS/123-QED}

\title{A Thermodynamic Constraint for the Electronic Structure of Fe-Ni Alloys at Room And High-Temperature}

\author{\textit{Jonathan Paras, Antoine Allanore}}
\affiliation{Department of Materials Science and Engineering,\\Massachusetts Institute of Technology, Cambridge, MA, USA }

\date{\today}

\begin{abstract}
The measurement of the electronic structure of metal alloys is hampered by the in-applicability of conventional quantum oscillation measurements owing to electron scattering by thermal or alloy disorder.  Recent advancements in the study of the electronic contribution to the entropy suggest a path to ground transport properties in alloy equilibrium thermodynamics.  Fe-Ni represents an interesting test-case because it exhibits an intricate electronic structure, order-disorder transformations,  and a large solid-solution region where equilibrium data can be obtained.  Using cluster modeling, the electronic contribution to the entropy can be inferred from high-temperature thermodynamic data.  Electronic transport property measurements can be used to independently evaluate the electronic contribution to the entropy.  Reconciling these two approaches at high temperature supports this method to study electronic structure for metal alloys.
\end{abstract}

\maketitle

\section{Introduction}

Before the advent of computerized electronic structure calculation methods,  the conventional method to determine the state of electrons in metals and alloys was to conduct experimental magnetic and electronic transport measurements\cite{nf1936theory,ashcroft1978solid,ziman1979principles}.
Such measurements,  like the Hall and De Haas van Alphen effects,  rely on magnetic field induced motion of electrons across their constant energy surfaces which enables the deduction of their available states.  Other transport properties like the thermopower,  Soret, and Nernst effects have thermodynamic definitions that connect quantum mechanical properties to measurable integral quantities like the Gibbs energy,  enthalpy,  and entropy \cite{Rockwood1984,rinzler2017thermodynamic,paras2020electronic,paras2021contribution}. 

There is therefore an opportunity to rationalize experimental thermodynamic properties and their models in the context of an electronic structure capable of reproducing the observed charge transport properties.  As Hurd points out,  however,  the complexities in defining energy-momentum surfaces in metals and alloys becomes manifold at high-temperature.  Simplifying assumptions to reproduce trends in transport properties variation with temperature and chemical composition are needed,  while avoiding the introduction of fudge-factors he calls "an exercise in numerology" \cite{hurd2012Hall,ziman1961ordinary}. 

The inability to reach the high-field condition in pure elements,  let alone alloys,  renders quantum oscillation measurements unable to support the study of the electronic structure of metallic systems as they approach melting and beyond. While methods like Angle Resolved Photoemission (ARPES) and x-ray photoemission spectroscopy (XPS) also probe the electronic structure and can be used at high-temperature, these techniques are sensitive to the electronic states on the surface of a material and unlike quantum oscillation measurements, may not be indicative of the bulk electronic states of the system. We therefore confine our comparison only to quantum oscillation measurements when discussing limitations in the prior art.

Our recent study of Cu-Ni \cite{paras2024evidence} showed the application of an irreversible thermodynamic formalism that links the thermodynamic definition of transport properties to the electronic structure. It suggested an opportunity to use thermodynamic integral quantities,  like the entropy and enthalpy,  to constraint the electronic structure of metals and alloys.  This may become a new method for high-temperature study of metallic systems.

\begin{figure}[h]
				\includegraphics[width=\linewidth]{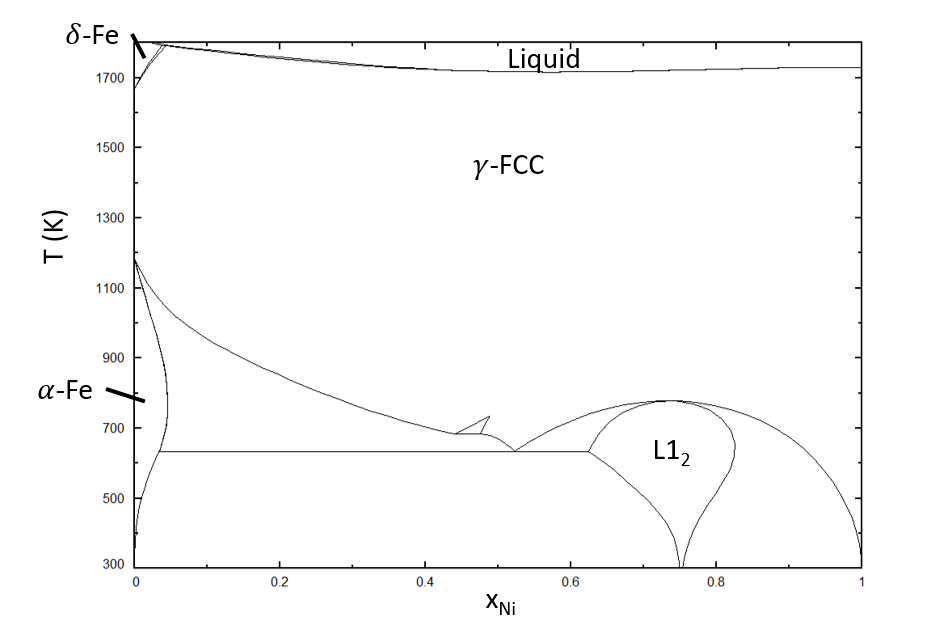}
				\caption{ Fe-Ni phase diagram,  computed using FactSage 8.0 using the FStel database \cite{Sundman1998}. The diagram illustrates the order-disorder transition of one low-temperature intermetallic compound to a high-temperature disordered solid solution (L1$_{2}$ to $\gamma$-FCC).}
    \label{FeNiPhaseDiagram}
				\end{figure}

For the study of Cu-Ni,  a simple one-band assumption worked satisfactorily in reproducing trends in the electronic entropy variation with composition,  as derived from low-temperature calorimetric measurements.  An example of the link with a thermodynamic solution model with cluster-configurational entropy was provided, qualitatively reproducing the trends in the thermodynamic entropy of mixing,  as well as dynamic properties of the melt \cite{paras2024evidence}.  Herein,  we pursue the application of the formalism for the Fe-Ni system,  which phase diagram at atmospheric pressure is shown in Figure \ref{FeNiPhaseDiagram} . 

Unlike Cu-Ni,  the Hall-effect and thermopower values for Fe-Ni change sign as a function of composition and phase,  possibly due to a multi-band electronic structure for at least some parts of the binary \cite{berndt1978electronic,hurd2012Hall,pugh1953Hall,pugh1955band}. A table for the values of the electronic contribution to the low-temperature heat capacity, also tabulated by Hultgren, is presented in Table (\ref{FeNiCalorimetricEentropy}). The most important feature is a local maximum of this entropy near the 30at.$\%$Ni concentration. This is very close to the INVAR composition, implying interplay between the electronic entropy and the localized magnetic and thermal expansion properties that have been observed for this composition.  These results indicate a large contribution to the entropy of mixing from the electronic states of Fe and Ni. 

As shown in Figure \ref{FeNiPhaseDiagram},  Fe-Ni alloys undergo BCC to FCC phase-transitions over a large composition range,  driven by the allotropic transition from BCC to FFC of pure Fe.  Low-temperature calorimetric electronic entropies are no longer meaningful anchors to study the high-temperature electronic entropy contribution without a better insight into Fe entropy across such allotropic transition.  Similar transitions are found for alloys with Zr, Ti,  and Co.  They all exhibit allotropic transitions before melting and frequently form multiple phases with large solubility with other elements in these various crystal structures.  This is a known issue when evaluating thermodynamic phase stability and solution models.  A good example of this is Al,  which only forms an FCC up to melting at 1 atm of pressure.  Al under both equilibrium and non-equilibrium conditions can form BCC structured alloys with elements including (but not limited to) Mg, Zr, Ti, Ni ,Sc, Li, and Hf \cite{malek2000structure,kobayashi1987new,wang2020thermodynamic,choudhuri2022investigation}. 

Therefore a new framework for understanding the high-temperature thermodynamic behavior needs to be proposed to interpret electronic entropy estimated from the transport properties. This framework will need to allow for multiple types of carriers. Fe-Ni will be our system of choice owing to its large solid-solution, order-disorder transformation and the existence of an allotropic transition in Fe.

Previous work has found that the relative importance of the vibrational entropy is greater for allotropic transitions (\cite{paras2020electronic,paras2021contribution}) than for alloy mixing (\cite{paras2024evidence}). Therefore, while the electronic contribution is the focus for analysis of solid solution mixing, the vibrational contribution will be examined in the context of the allotropic transition. We do however acknowledge that there is work that may suggest a large role for the vibrational entropy, and that efforts are ongoing to characterize this contribution \cite{li2020ground}. Because our arm is to link the electronic contribution to cluster configurational and EMF thermodynamics, rather than compute phase diagrams, it is not within the objective of our current work to address the vibrational contribution in solid solutions.

%
%
%

Elsewhere,  solution thermodynamic modeling of metal alloys are frequently revised and re-optimized to reproduce experimental thermodynamic data as well as phase stability domains \cite{harvey2020elaboration,lupis1967prediction,lehmann2010generalized}.  Intricate cluster-based solution models of Fe-Ni are computationally very tractable,  supporting a reinterpretation of the existing thermodynamic data,  as detailed in Appendix A.  It is experimentally reported that while the entropy of mixing in Fe-Ni alloys seems nearly ideal,  the enthalpy of mixing is reportedly large,  implying that there must be significant ordering in the system,  itself indicative of a non-ideal configurational entropy.  


The Fe-Ni binary alloy also exhibits an intermetallic compound FeNi$_{3}$ which undergoes an order-disorder reaction at 776 K \cite{hultgren1973selected}. The enthalpy of this phase transformation was reported to be $\delta H^{T}$ =  2719 J/mol resulting in an entropy of disordering of about $\delta S^{T}$ = 3.5 J/mol K. The review by Wakelin and Yates in 1953 demonstrates that the ordering is kinetically limited,  often resulting in highly hysteretic behavior wherein the order-disorder transformation temperature collides with the Curie temperature near 850 K \cite{hultgren1973selected,wakelin1953study}. Whereas experimental evidence exists for the ordering of FeNi$_{3}$,  some phase diagrams include FeNi and Fe$_{3}$Ni; they exhibit slower ordering kinetics and the weak evidence of their existence will preclude their integration in this study. 

Herein we report experimental measurements of the resistivity and thermopower of Fe-Ni binary alloys across the composition and temperature space,  including the high-temperature solid solution up to 1200K.  We interpret the results in the context of the recently proposed formalism, discussing role of the allotropic phase transition in Fe and the disordering of FeNi$_{3}$.

Combined with the Hall-effect data from the litterature,  the findings are used as input to the electronic state entropy of the alloy at both low and high temperature.  Independently,  we fit the central atom model - a cluster-based,  quasi-chemical solution model - to the integral thermodynamic data available at 1200 K,  to evaluate a non-configurational excess entropy.  We then evaluate what electronic properties (charge carrier types and concentration) would be needed in order to support the attribution of this excess entropy to an electronic entropy as evaluated via the formalism devised for Cu-Ni. \cite{harvey2020elaboration,paras2020electronic,paras2024evidence}.

\begin{table*}[!htb]
    \centering
     \renewcommand{\arraystretch}{1.2}
    \begin{tabular}{cccccccccccccc}

        \toprule
        xNi & 1       & 0.894 & 0.777 & 0.69 & 0.587 & 0.492 & 0.447 & 0.389 & 0.338 & 0.292 & 0.190 & 0.093 & 0  \\
        \midrule
         $\gamma  \text{(J/mol K}^{2}\text{)} \times 10^{4}$    & 72.8 & 59.8    & 43.9  & 42.2 & 42       & 47.69  & 54.81  & 74.1     & 91.6    & 139.3 & 62.9   & 61.9       & 50.2 \\
        \bottomrule
    \end{tabular}
    \caption{Low temperature electronic heat capacity data for Fe-Ni alloys compiled by \cite{hultgren1973selected}.}
    \label{FeNiCalorimetricEentropy}
\end{table*}

\section{Methods}

Samples of Fe-Ni alloys were prepared in-house using a Buehler AM-500 arc-melter by combining pure chunks of Fe, Ni. Total sample button mass was consistently 50 g. Electrolytic grade Fe (99.95$\%$) was sourced from TOHO ZINC Corporation in chunk format while slugs of Ni were sourced from Thermo Fischer (Puratronic, 99.99$\%$). Samples were remelted 5 times within the arc-melter under Ar cover gas after the system was purged using a diffusion pump to 10$^{-6}$ mbar.  Between each melt, a sacrificial titanium getter was remelted and allowed to fully solidify.  The sample crucible was made of Cu and actively water cooled. Samples were consistently flipped for each remelting to ensure button homogeneity.  Meaningful differences in composition were not observed by EDS observation of the alloy composition.  After alloying, ingots of roughly 60 mm x 20 mm x 8 mm were suction cast in an AM-500 arc-melter using identical operating conditions. Suction casting was noted to contribute to a decrease in segregation in the alloyed samples.  X-ray diffraction (XRD,  patterned show in Appendix B) was conducted to verify that samples were single-phase and to also determine the BCC to FCC transition for samples at room temperature. A Mo source was used due to the fluorescence of Fe under Cu irradiation (reported in Appendix B), however Cu K-$\alpha$ was also used to confirm the results.

\subsection{Rolling and Heat Treatment}

20 mm x 8 mm x 4 mm samples were cut using a wire EDM from the rectangular ingots produced from suction casting. Samples were then cold-rolled along the direction of the long axis to a minimum of 30$\%$ thickness reduction while cold using an N. Ferrara INC hand roller. Rolled samples were then loaded into quartz ampoules that were purged with Ar 3 times before being vacuum sealed at 10$^{-2}$ mbar. A special Zr-V-Fe gettering alloy was used to clean any remaining oxygen from the ampoule during annealing (SEAS Getters). Homogenizing temperatures varied between 1050 and 1300\textdegree C. Annealing was conducted to remove the effects of cold rolling on the transport properties and coarsen the grains in order to decrease the grain-boundary effect on subsequent measurement.

Machining for final transport property specimen dimmensions was conducted in a Hansvedt traveling wire-EDM to a final dimmension of 22 mm x 3 mm x 3 mm. Samples were then hand polished on each face down to 1200 grit using SiC and Emery paper to ensure quality electrical contact during transport measurements. This final step proved crucial to ensure the reliability of electrical contact, particularly at high-temperature.

\subsection{Transport Measurements}

An ULVAC-RIKO ZEM-3 was used to measure both the electrical resistivity and the thermopower as a function of temperature. The furnace was in high-temperature configuration, so measurements were conducted between room temperature and 1000$^\circ$C. The system was calibrated using a Constantan calibration sample (provided by ULVAC) of identical geometry to those used in this study. Care was taken to gently lap the type-R thermocouple probes using emery or SiC paper (800-1200 grit) before and after experiments to ensure quality contact during the course of measurements. Measurements were automated using ULVAC software and were conducted under He gas after He purging of the chamber.  He gas was scrubbed using a Ti gettering system from OxyGon (OG-120M) achieving $<$ 50 ppb O$_{2}$ of the outlet process gas (according to the built-in oxygen sensor). 

Each sample was measured 3 times at each temperature point, and at-least two samples were used for each experimental composition. The results were then averaged together for a composite curve.  Error associated with the ZEM-3 measurement geometry was quoted by the manufacturer to be 7$\%$ for thermopower measurements and 10$\%$ for resistivity. Deviation among samples tested never exceeded these limits, therefore the manufacturer quoted error percentages should be used when considering expected experimental error for each data-point. The error bars on the figures will otherwise be suppressed for reading clarity.

\subsection{Entropy of FeNi$_3$ order-disorder transformation}

The Fe-Ni phase diagram is often reported with up to three intermetallic compounds, Fe$_{3}$Ni, FeNi and FeNi$_{3}$. The sluggish nature of the formation of the first two has relegated their study largely to computational efforts and the rare meteorite sample that would nominally have allowed for long ordering times \cite{dang1996simultaneous,albertsen1978mossbauer}. While still hysteretic, transport properties measurements of FeNi$_{3}$ like resistivity \cite{wakelin1953study} and spectroscopic studies like Mossbauer methods have been implemented to track the order-disorder transformation of the 75 at.$\%$ Ni compound. Hultgren compiled and reviewed the thermodynamic available thermodynamic data that attempted to measure the total enthalpy change, and by extension, the entropy change for the disordering of this compound into the solid solution\cite{hultgren1973selected}. The consensus value for the entropy change was found to be closer to that of Ida and Kubaschewski,  at about 3.5 J/mol K \cite{kubaschewski1967heats}. There does not exist Hall effect data for FeNi$_{3}$ at high-temperature as a function of disordering,  nor low-temperature data for the fully ordered state. Therefore, only room-temperature values of Galepov are used to calculate the electronic entropy change; this will render the conclusions qualitative in nature as previous work has demonstrated that the Hall-effect can change significantly as a function of the degree of atomic ordering \cite{galepov1969Hall,paras2021contribution,Elkholy1962}. 

\subsection{Central Atom Model}

A cluster model,  the Central Atom Model,  as derived in C.H.P. Lupis \cite{lupis1983chemical} will be used to derive an excess entropy for Fe-Ni alloys at 1200K.  A detailed account of the modeling steps are given in Appendix A.  The steps are summarized as follows: 

\begin{itemize}
    \item A next-neighbor arrangement centered on either a Fe or Ni atom is formulated for the solid solution alloy,  each cluster is described as inherited from a quasi chemical reaction of the corresponding Fe and Ni atoms
    \item Experimental enthalpy of mixing data are used to fit the corresponding energetic (bond) parameters and identify the maximum probability of distributions of clusters
    \item This generates a cluster configurational entropy of mixing that is consistent with the experimentally measured enthalpy of mixing
    \item Subtracting the corresponding configurational entropy to the experimentally measured entropy of mixing leads to an excess entropy,  which here we ascribe to the electronic entropy
\end{itemize}

While the corresponding enthalpy and entropy may be used to generate phase boundaries,  here instead we evaluate what would be the electronic entropy necessary to reproduce such excess entropy,  given the available transport measurements and a multi-carrier interpretation of the electronic entropy. 

\subsection{Interpretations of the Electronic Entropy in Multi-carrier Systems}

Significant work has been conducted to rationalize the conflict between electronic transport measurements like the thermopower and Hall effect, that conflict on the predominant carrier type.  Such efforts are well summarized by Hurd but originally presented by Chambers \cite{hurd2012Hall,chambers1952two}.  As pointed out in \cite{paras2021contribution},  if there are two bands near the Fermi-level, and scattering between those bands is largely neglected so that they behave as if they were parallel conduction channels, the thermopower would assume the form 

					\begin{eqnarray}
					\ \alpha = \frac{\sigma_{n} \alpha_{n}+\sigma_{p} \alpha_{p}}{\sigma_{n}+\sigma_{p}}
     \label{ seebeckMultiband}
					\end{eqnarray}
Where $\alpha_{i}$ and $\sigma_{i}$ ($i$ indice is $n$ or $p$) are the partial thermopower and conductivity of those respective electron- and hole-like bands. This results in an electronic entropy that is given by 
					\begin{eqnarray}
					 S_{e}=-n_{n}e\alpha_{n}+n_{p}e\alpha_{p}
					\label{MultibandEntropy}
                        \end{eqnarray}
This requires knowledge of the partial band conductivity as well as the entropy per carrier in those bands.  Furthermore, interpretation of the Hall coefficient would need to include electron- and hole-like contributions.  Seminal work by Takano demonstrated that such disagreements between the Hall and thermopower can be rationalized by considering regions of the Fermi surface to have both positive and negative curvature\cite{takano1967interpretation}. His demonstrations used a simple model that parameterized these regions of curvature as two effective bands, resulting in Equation (\ref{MultibandHallEffect})

					\begin{eqnarray}
			 R_{H}=-\frac{1}{(|e|N_{o}(\sigma_{e}+\sigma_{h})^{2})}\left (\frac{\sigma_{e}^{2}}{n_{e}}-\frac{\sigma_{h}^{2}}{n_{h}} \right ) 
    \label{MultibandHallEffect}
					\end{eqnarray}
					
Where $\sigma_{i}$ are interpreted in terms of the Drude model.  However, this would result in more unknowns than equations, as the electron mobility of individual bands would need to be computed,  or backed out through new experimental methods.  Applying a relatively more sophisticated treatment to the evaluation from transport data in line with the proposed equations is the occupation of ongoing work. This would then allow the electronic entropy of materials that exhibit multiple sign changes of their thermopowers and Hall effects to be evaluated and compared to experimental thermochemical data. The details of such a multi-carrier model have been discussed in prior work on the matter for Cu-Ni alloys \cite{paras2024evidence}.

Equation (\ref{MultibandHallEffect}) is what is refereed to as the low-field limit, where the Hall response is a mixture of the density of electrons at the Fermi-surface as well as their scattering behavior.  This is the expected regime for measurements of polycrystalline samples and the "relatively" high temperature (300K).

The formalism used herein links such electronic properties and the electronic state entropy.  To calculate the state electronic entropy used to fit the sub-band thermopower and electronic distributions,  we evaluate the state electronic entropy at any temperature via equation 

\begin{equation}
    S^{e} = (1-x_{\text{Ni}})S^{e}_{\text{Fe}} + x_{\text{Ni}}S^{e}_{\text{Ni}} + \Delta S^{e}_{\text{mix}}(x_{\text{Ni}})
    \label{StateEEntropyThermo}
\end{equation}

The term $S^{e}_{\text{Fe}}$,  electronic entropy of pure Fe was evaluated from low-temperature electronic heat capacity with a correction added from the change in electronic entropy for BCC ($\alpha$)$\rightarrow$FCC Fe ($\gamma$) (see next section).  The term $S^{e}_{\text{Ni}}$ was evaluated from the single-band assumption for pure Ni at 1200 K already presented in the previous work on Cu-Ni \cite{paras2024evidence}. The term $\Delta S^{e}_{\text{mix}}$ is obtained via the experimental charge transport properties measurements of the alloys as described above.

\subsection{Allotropic Transition of Fe}

The cluster model produces an excess entropy of mixing that we ascribe to the electronic entropy of mixing.  However,  Equation \ref{MultibandEntropy} is a total state electronic entropy,  from which to anchor a multi-carrier interpretation of the electronic structure of the Fe-Ni alloy at high temperature can be proposed.  This means the electronic state entropy of pure Ni and pure Fe are needed,  at both low and high temperature,  to evaluate the state entropy from the mixing entropy.

 Nickel charge transport properties at low temperature are well described by a single-type of charge (electron) and it undergoes no allotropic transition such that the electronic entropy of pure Ni is available, as discussed and presented in \cite{paras2024evidence}.  On the contrary,  Fe undergoes several allotropic transitions,  the one of relevance in this study being the $\alpha$ $\rightarrow$ $\gamma$. Therefore,  we propose to estimate the electronic contribution of this transition in order to correctly project the low-temperature heat capacity of BCC iron to the high temperature FCC regime.  This is not an easy task, due to iron complicated Fermi-surface. The low-temperature calorimetric measurements in the Fe-rich portion of Fe-Ni alloys cannot be used either,  as they also undergo a BCC to FCC transition.  

The partitioning of the entropy of a phase transformation into its various components is commonly written:
					\begin{eqnarray}
					\delta S^{Total}= \delta S^{vib} + \delta S^{mag} + \delta S ^{elec}
     	\label{totalentropy}
					\end{eqnarray}
where the terms on the right hand side corresponds to the vibrational, magnetic,  and electronic entropy changes across the transition.  Table \ref{AllotropicTable} provides the litterature data used herein to further evaluate the possible contribution of mobile electron/holes,  $\delta S ^{elec}$.

At atmospheric pressue,  Fe exhibits several solid-state phase transitions,  first demagnetizing at $1043 K$,  then experiencing the allotropic transitions ($\alpha$) to ($\gamma$) at T$^{\alpha\rightarrow\gamma} = 1185 K$,  and reverting to BCC ($\delta$) at T$^{\gamma\rightarrow\delta} = 1667 K$.  The total entropies of the allotropic transitions have been measured calorimetrically and compiled by Hultgren. They were found to be 0.79 and 0.67 J/mol K respectively \cite{hultgren1973selected}. Contemporary measurements reported by Neuhaus et al.  put these values closer to 0.76 and 0.50 J/mol K respectively \cite{Neuhaus2014}. 

Direct measurement of the phonon dispersion relations in $\alpha$,  $\gamma$,  and $\delta$ Fe,  have been proposed,  from which estimates of the vibrational entropy can be deduced \cite{Neuhaus2014}.  It suggests a vibrational entropy $\delta S^{\alpha\rightarrow\gamma}_{vib} =0.32$ and  $\delta S^{\gamma\rightarrow\delta}_{vib} =0.46$ J/mol K.  This indicates that 50$\%$ of the $\alpha\rightarrow\gamma$ phase transition entropy remains unaccounted for.  Because the ferromagnetic transition in iron is not concomitant with the relevant structural transition,  we hypothesize that this discrepancy can be accredited to the electronic entropy alone.  We therefore use the experimentally measured charge transport properties of Fe to evaluate the contribution of electrons to iron entropy across the ($\alpha$)$\rightarrow$($\gamma$) transformation.  We assume that the number of free carriers do not vary significantly across the phase transitions of interest in  Fe, and that the nature of the bonding does not change significantly across the structural transition of this high-density solid.  We provide similar insight for Co,  as analogies, which also undergo an allotropic transformation without second order magnetic changes, though between different crystal structure (HCP to BCC).

\section{Results}

	\subsection{Electronic Entropy Change for Iron BCC$\rightarrow$FCC } 

\begin{figure}[h]
    \includegraphics[width=\linewidth]{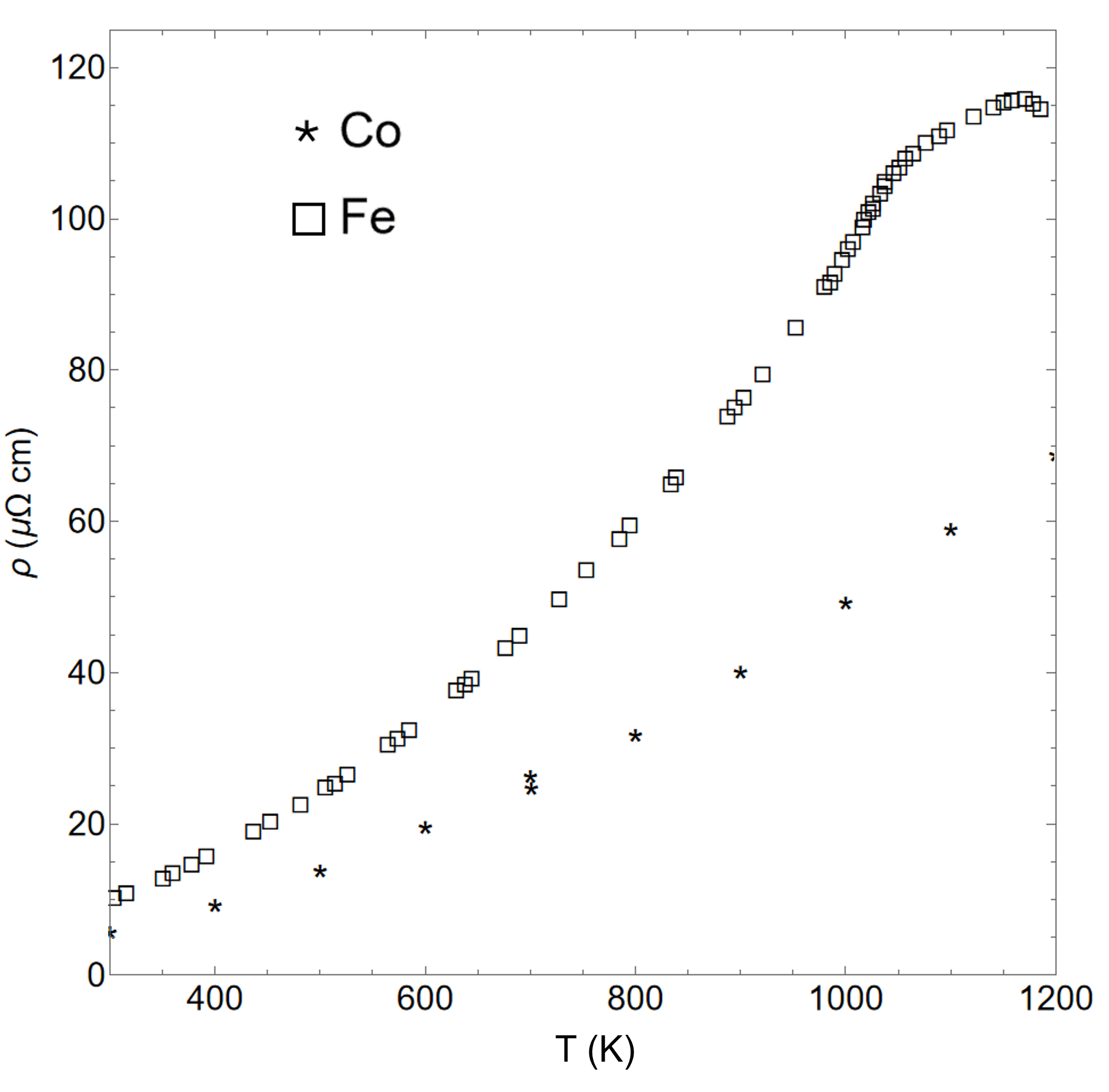}
    \caption{Temperature variation of the resistivity for Co and Fe, replotted from \cite{Betteridge1980,arajs1971thermoelectric}.}
    \label{cobaltironfig}
\end{figure}

The resistivity of Fe through its BCC$\rightarrow$FCC transformation is shown in Figure (\ref{cobaltironfig}),  along the one for Co (HCP$\rightarrow$BCC) as analogy.  No large discontinuities are observed through the allotropic transition.  This contrasts with other allotropic phase transformations for other metals,  such as Sn,  where instead the resistivity increases several orders of magnitude upon forming the low temperature $\alpha$ phase. In the case of allotropic transitions that do not coincide with metal-insulator transitions, we do not expect a large change in resistivity. This implies in Fe an allotropic phase transformation for which we expect large changes in the free carrier concentration \cite{di2009monitoring}.  Indeed other phase transitions studied using Equation (\ref{MultibandHallEffect}) have demonstrated order of magnitude changes in the carrier concentration which drive large changes in the electronic entropy but this is unlikely considering the relatively small entropy change in allotropic transitions \cite{Cusack,Cutler1966,paras2020electronic,paras2021contribution}. 

Calculations of the change in electronic entropy equation (\ref{MultibandEntropy}) for the $\alpha\rightarrow\gamma$ transition in Fe are presented in Table (\ref{AllotropicTable}). We compare this with contemporaneous estimates of the vibrational entropy.  The thermopower data were taken from Fulkerson and Vedernikov\cite{Fulkerson1968,Vedernikov1969}.

\begin{table*}
\begin{ruledtabular}
\begin{tabular}{cccccc}
Material & $T_{c}$ (K) & $\Delta S^{elec}$ (J/molK) & $\Delta S^{vib}$ (J/molK) & $\Delta S^{Total}_{calc}$ (J/molK) & $\Delta S^{Total}_{exp}$ (J/molK)
 \\[0.2pt]
 \hline 
 \\[0.2pt]
 Fe & 1185 & 0.30 & 0.32\cite{Neuhaus2014} & 0.62 & 0.76$\rightarrow$0.79 \cite{hultgren1973selected,Neuhaus2014}
 \label{AllotropicTable}
\end{tabular}
\end{ruledtabular}
\caption{Litterature and calculated data for the entropy Fe allotropic transformation from BCC to FCC.  The electronic contribution calculated herein from experimental data represents approximately 50\%~of the total entropy change.}
\end{table*}

	\subsection{Transport Properties of Fe-Ni Alloys}
	
\begin{figure*}[!htb]
\label{FeNiResistivty}
				\includegraphics[width=\textwidth]{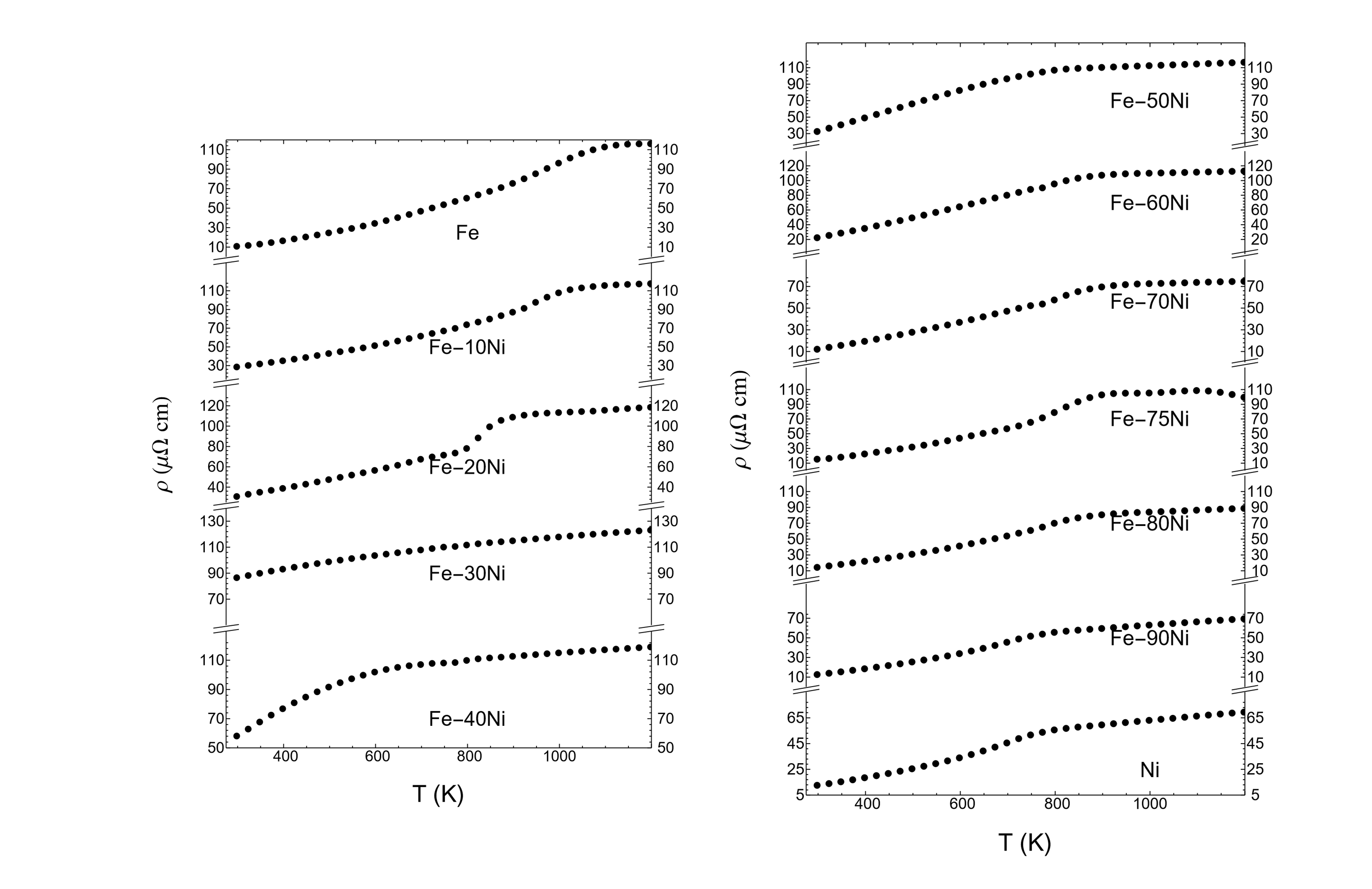}
				\caption{The resistivity was measured in the Fe-Ni alloy system as a function of temperature. The temperature and composition dependence are plotted here. Note the discontinuous y-axis.}
    \label{FeNiResistivity}
				\end{figure*}

The thermoelectric power and resistivity of Fe-Ni alloys measured herein are presented in Figures \ref{FeNiThermopower} and  \ref{FeNiResistivity} respectively.   The trend with composition at room-temperature shows increasingly negative thermopower as Fe is added to the system until roughly 50$\%$ Ni,  where a sharp rise to near zero is measured at 30$\%$,  coinciding quite close with the Invar composition of this alloy.  The BCC to FCC transition, confirmed with the XRD characterization (Appendix B) is marked by a large change in the resistivity,  particularly for the Fe-10Ni and Fe-20Ni samples,  which are BCC at room temperature and undergo a martensitic transformation to the $\gamma$(FCC)-phase.  The rest of the composition range crystallized in the FCC structure after solidification.

\begin{figure}[h]
                
                \centering
				\includegraphics[width=1.0\linewidth]{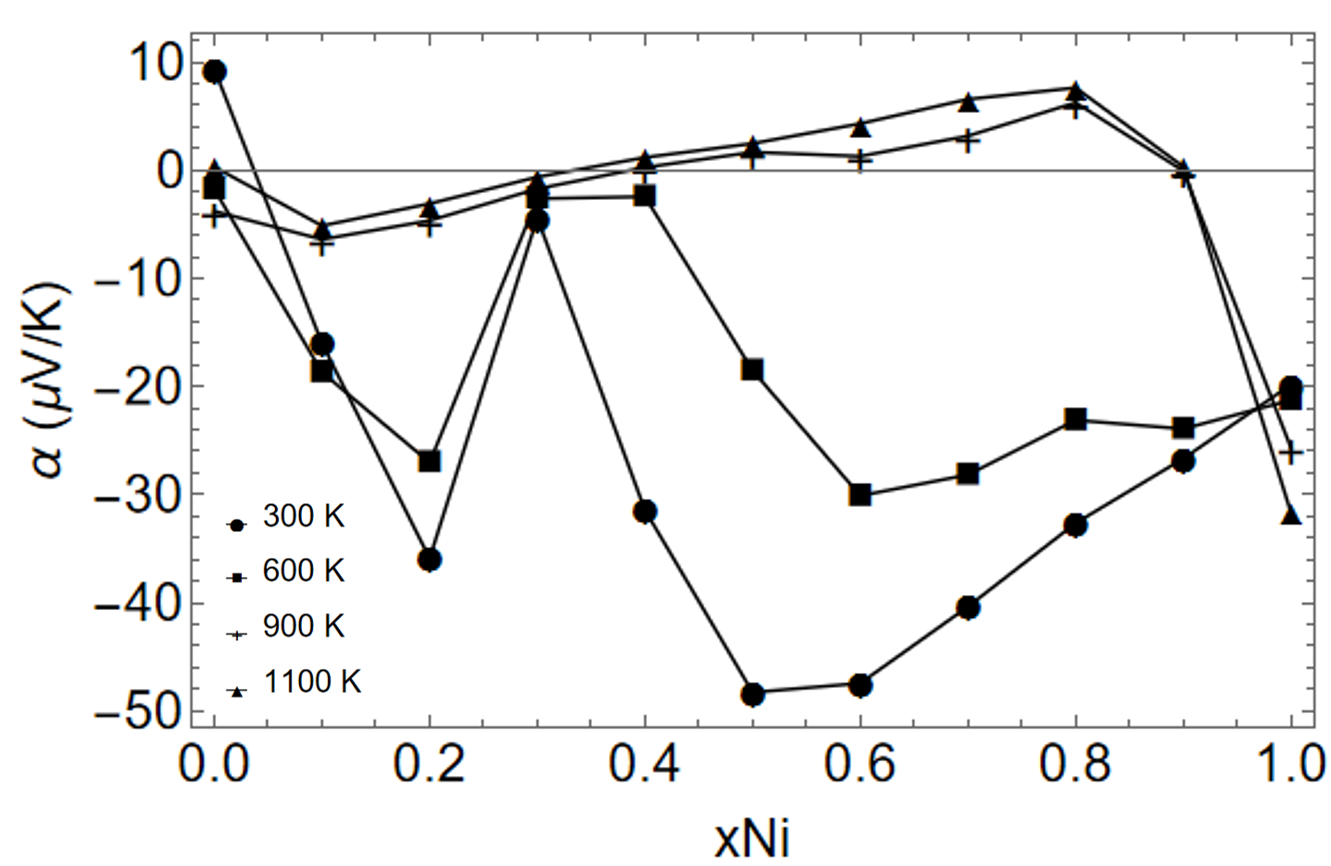}
				\caption{Isothermal plots of the experimental thermopower in Fe-Ni alloys.}
				\label{FeNiIsothermalThermopower}
				\end{figure}

\begin{figure}[h]
                
                \centering
				\includegraphics[width=1.0\linewidth]{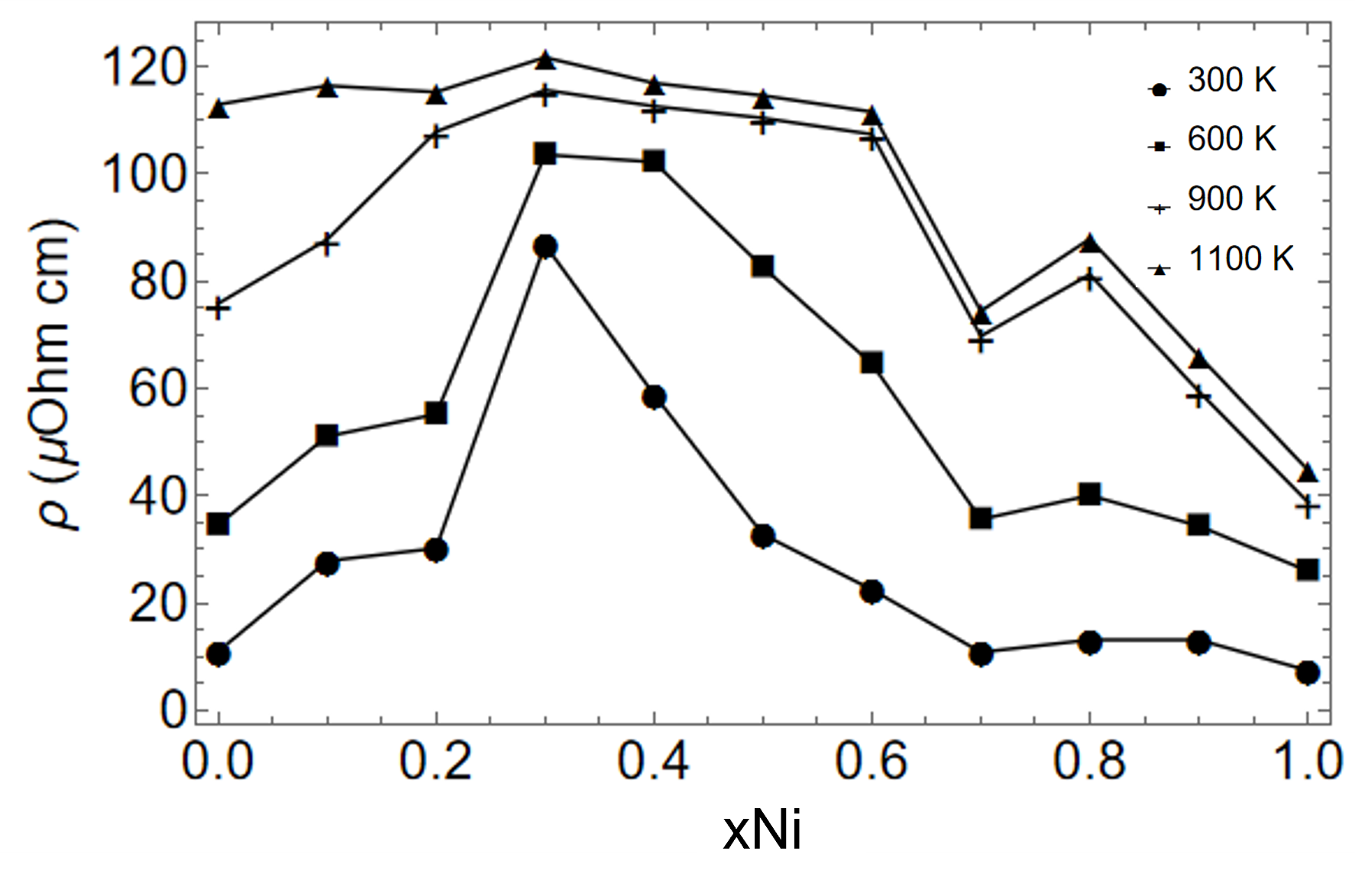}
				\caption{Isothermal plots of the experimental resistivity in Fe-Ni alloys.}
				\label{FeNiIsothermalRes}
				\end{figure}

\begin{figure}[h]
                
                \centering
				\includegraphics[width=1\linewidth]{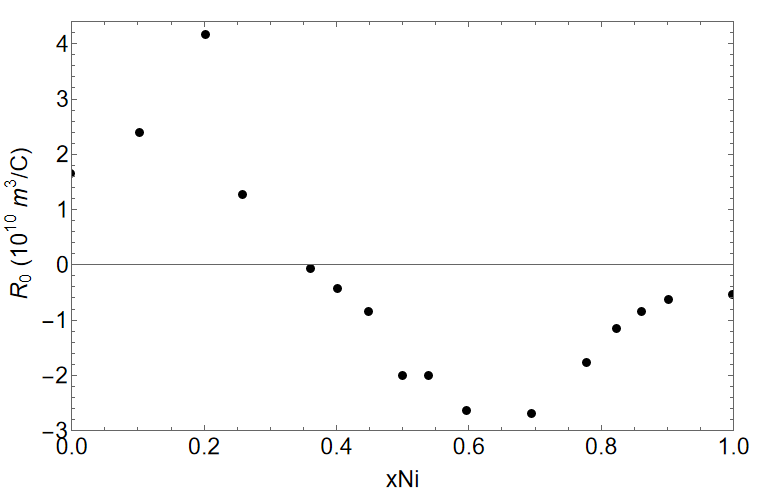}
				\caption{Isothermal plots of the Hall-effect measured by Galepov \cite{galepov1969Hall}. Note the change in sign that occurs near 30at.$\%$Ni as indicative in a change in the dominant carrier type.}
				\label{FeNiHallGalepov}
				\end{figure}

Isothermal plots of the thermopower of Fe-Ni alloys are given in Figure (\ref{FeNiIsothermalThermopower}), resistivity in Figure (\ref{FeNiIsothermalRes}) and the Hall-effect in Figure (\ref{FeNiHallGalepov}).

These Figures highlight the existence of a local-maximum in the resistivity near Fe-30Ni, local minimum in the thermopower and Hall-effect, and a local-maximum in the calorimetrically derived electronic entropy.

\begin{figure*}[!htb]
\label{FeNiThermopower}
				\includegraphics[width=\textwidth]{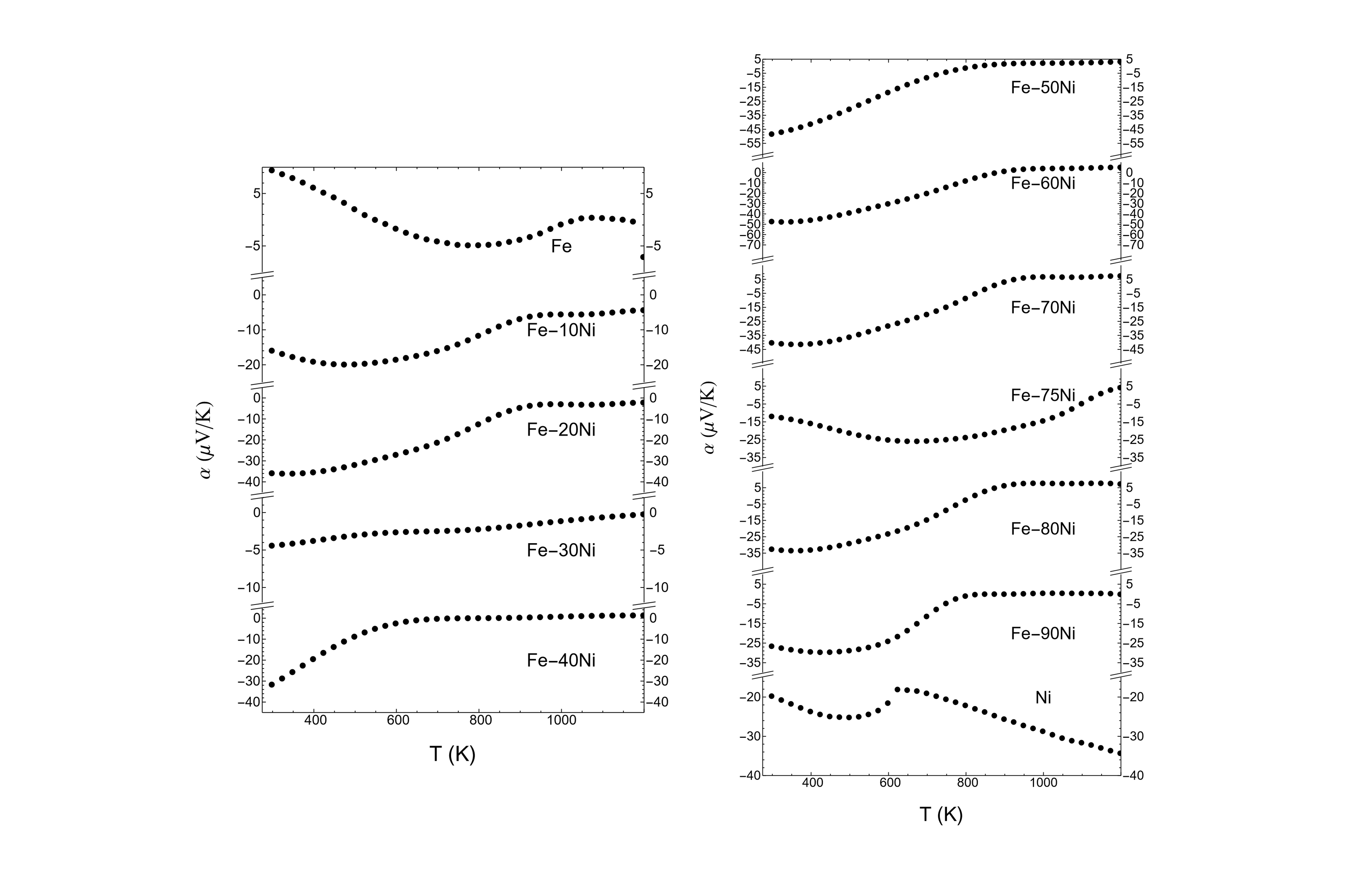}
				\caption{The absolute thermopower was measured in the Fe-Ni alloy system as a function of temperature. The temperature and composition dependence are plotted here. Note the discontinuous y-axis.}
    \label{FeNiThermopower}
				\end{figure*}

	\subsection{Central Atom Model of Fe-Ni Alloys}

The fitting values were produced using a non-linear model fit on the data from Ono et al at 1200 K\cite{ono1977thermodynamic},  as detailed in Appendix A.  Mathematica was used with using the Levenberg-Marquardt method via the parameter space was constrained to physical values of the bonding parameters and cluster stoichiometry as: 

\[
\begin{aligned}
3 > \alpha_1 > 0, & \quad 3 > \alpha_2 > 0, \\
-3 < \alpha_3 < 0, & \quad -3 < \alpha_4 < 0, \\
-3 < \alpha_5 < 0, & \quad -3 < \alpha_6 < 0, \\
& \delta\psi_{1\text{B}}^\text{A}  < 0
&
& \delta\psi_{1\text{B}}^\text{B}  < 0
\end{aligned}
\]

Figure (\ref{fig:CAMFit}) shows the enthalpy of mixing from the Central Atom Model with the parameters listed in  Appendix A, Table (\ref{model_params}).  It reproduces key trends with composition of the enthalpy of mixing data of Ono\cite{ono1977thermodynamic} at 1200K.  The Fe-Ni bond energy is negative,  and a driving force is observed for the formation of the FeNi$_{3}$ compound that is asymmetric but otherwise negative.

\begin{figure} 
    \centering
    \includegraphics[width=\columnwidth]{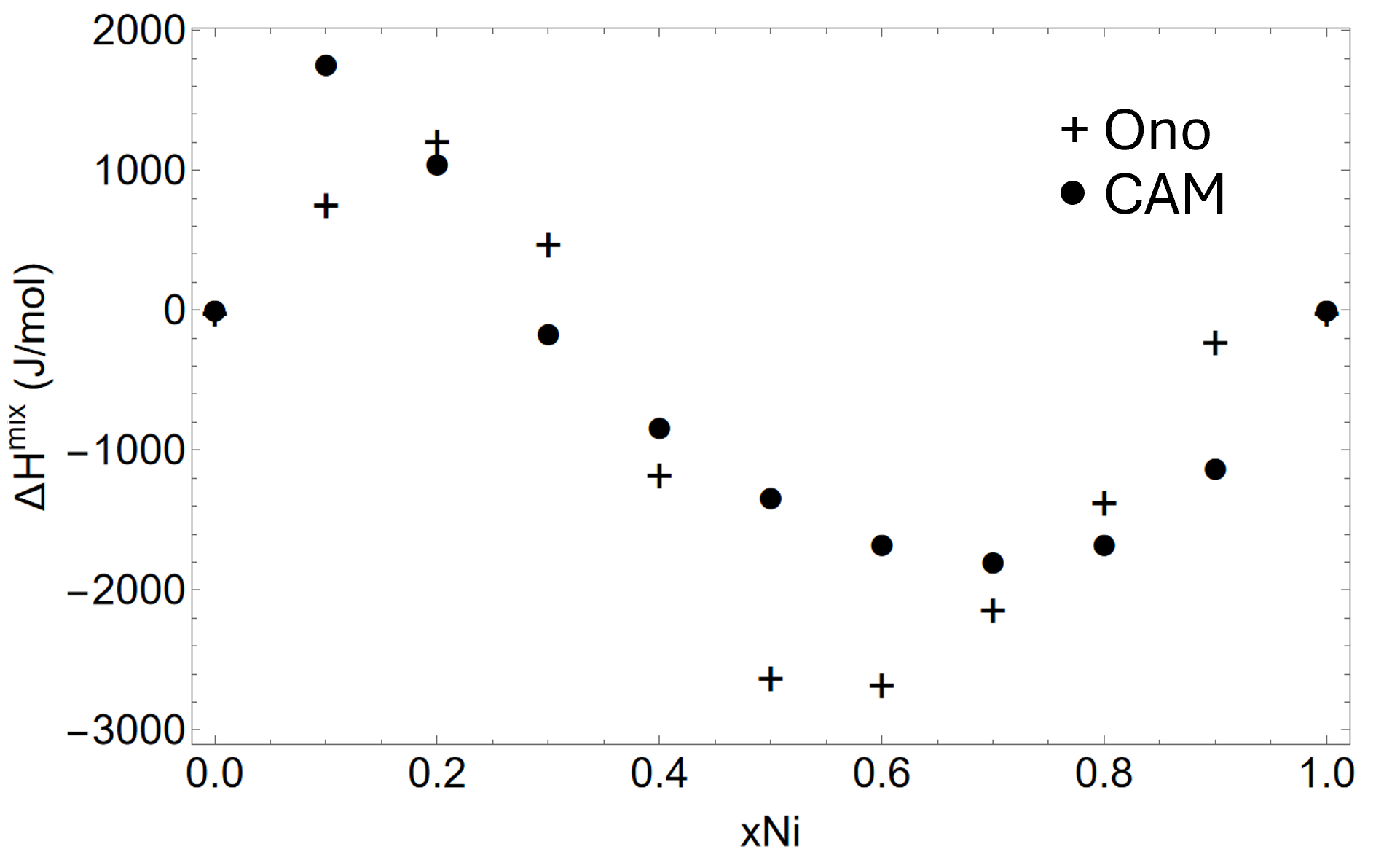}
    \caption{Enthalpy of mixing data of Ono and the CAM model fit compared}
    \label{fig:CAMFit}
\end{figure}

\begin{figure}[h] 
    \centering  \includegraphics[width=\columnwidth]{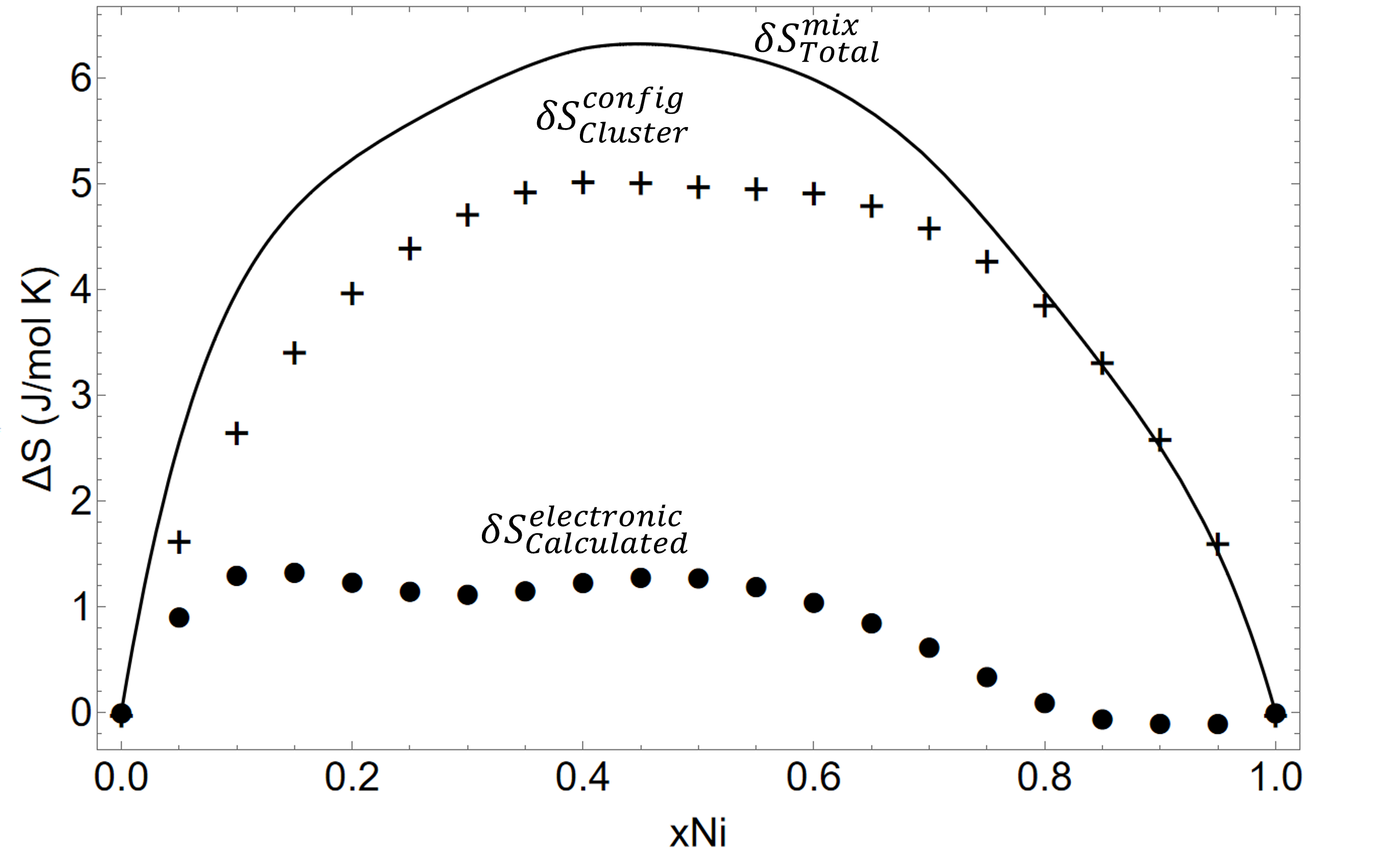}
    \caption{The total experimental entropy of mixing is plotted along with the cluster configurational entropy consistent with the enthalpy of mixing in Figure (\ref{fig:CAMFit}). The difference of these two curves results in an excess entropy which is ascribed to the electronic component.}
    \label{entropycomparison}
\end{figure}

The corresponding configurational entropy due to the random mixing of those clusters is compared with the total measured entropy of mixing from Ono in Figure (\ref{entropycomparison}).  It is clear,  as expected for such model,  that there is a non-configurational entropy contribution that is not accounted for.  This missing (``excess'') entropy is herein ascribed to the mobile electrons/holes at high-temperature.  This suggest a relatively large electronic entropy of mixing that is unaccounted for when the bonding energies of a cluster model are fit to a non-zero enthalpy of mixing.

	\subsection{Electronic Entropy of Fe-Ni Alloys}

			\subsubsection{Electronic Entropy at 300 K}

\begin{figure}[h]
            
                \centering
				\includegraphics[width=1.1\linewidth]{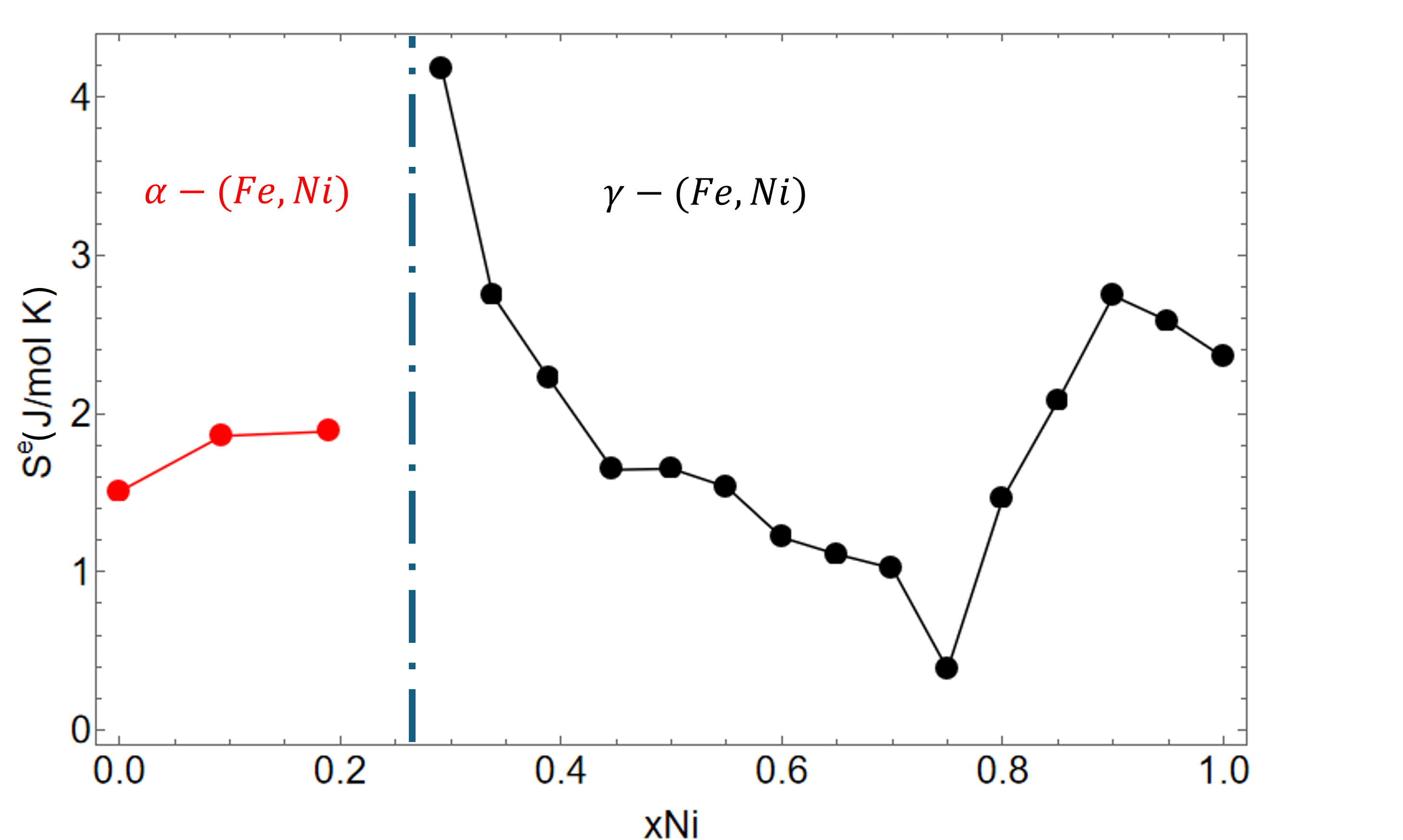}
				\caption{The electronic entropy,  at 300K,  for BCC (\textless 30~at.\% Ni) and FCC (\textgreater 30~at.\% Ni) Fe-Ni alloys based on a multi-carrier model for 0--50\% Ni and a single-band model from 50\% to 100\% Ni. Entropy derived from low temperature calorimetry data (Table \ref{FeNiCalorimetricEentropy}) and a naive one-band model are shown in Appendix D for comparison.}
\label{FeNielectronicentropyOneband}
				\end{figure}

The electronic entropy of Fe-Ni alloys were evaluated based on Hall-effect data from Galepov \cite{galepov1969Hall},  and the transport data reported in Figures  \ref{FeNiThermopower} and  \ref{FeNiResistivity}.  A two-band model from Equation (\ref{MultibandEntropy}) was then fitted to the results,  the results of which are shown in Figure \ref{FeNielectronicentropyOneband}.  The results are a quantitative account of the entropy per electron and hole as a function of composition, as well as the distribution of such electrons and holes shown in Appendix D

The comparison of the experimentally derived electronic entropy and the low-temperature entropy as derived from the electronic heat capacity. (Table (\ref{FeNiCalorimetricEentropy}) is plotted in Figure (\ref{FeNielectronicentropyOneband}).  Broad agreement between the electronic entropy of the single-band model and the calorimetric electronic entropy was observed for compositions greater than 50 atomic percent Ni.

The anchor values for pure Ni and pure Fe at room temperature were chosen based on De-Hass Van Alphen measurements discussed earlier to deduce the Fermi-surface of Fe, and it was presumed based on previous work that at such temperatures Ni behaved as if it were dominated by a single carrier type \cite{paras2024evidence,stearns1971origin,stearns1973origin,stearns1977simple,pugh1953Hall,pugh1955band}.

\subsubsection{Electronic Entropy of Fe-Ni Alloys at 1200 K}
The electronic state entropy was evaluated using two-band model fit to the measured transport data and the excess entropy of mixing results from the difference between the entropy of the Central Atom Model and the measured thermodynamic data of Ono \cite{ono1977thermodynamic,lupis1983chemical}.  Note that the electronic entropy estimated for the BCC to FCC transition in Fe from the previous section is used (0.3J/molK) to evaluate the electronic entropy of pure Fe at this temperature. 
		
The key assumption necessary to arrive at these results was to propose the electron distributions shown Appendix C.   In addition to the compensation point at 30at.$\%$Ni,  where we presumed the carrier numbers needed to meet,  we otherwise connected these "hanging" points with lines.  Small deviation from the proposed values resulted in non-physical values for the subband thermopowers, namely negative values for the hole contribution and positive values for the electron. This was used as a self-consistency check in the absence of additional constraining equations.

The values proposed for a full-two band model, as opposed to the single carrier assumption derived for the allotropic phase transition in Fe, use starting values for the number of electrons and holes per atom according to subsequent work by Stearns beyond the RKKY model whereby $n_{e} = 0.8$ and $n_{h} = 0.3$ (per atom) \cite{stearns1971origin,stearns1973origin,stearns1977simple}. This suggests that the electronic contribution to the entropy upon allotropic transition of Fe may involve different electrons than those that contribute to the mixing with Ni into a solid-solution. These values were then made to be equal at 30 at.$\%Ni$ and asymptoted to the single-electron assumption because of the congruity in the transport and calorimetric electronic entropy in this region. For the high-temperature value system, it was necessary to jump the number of holes because of the shift towards positive thermopower values with a logistic function as linear behavior was insufficient to reproduce self-consistent subband thermopowers with the proper sign. 
		
\begin{figure}[h]
    \centering
    \includegraphics[width=1\linewidth]{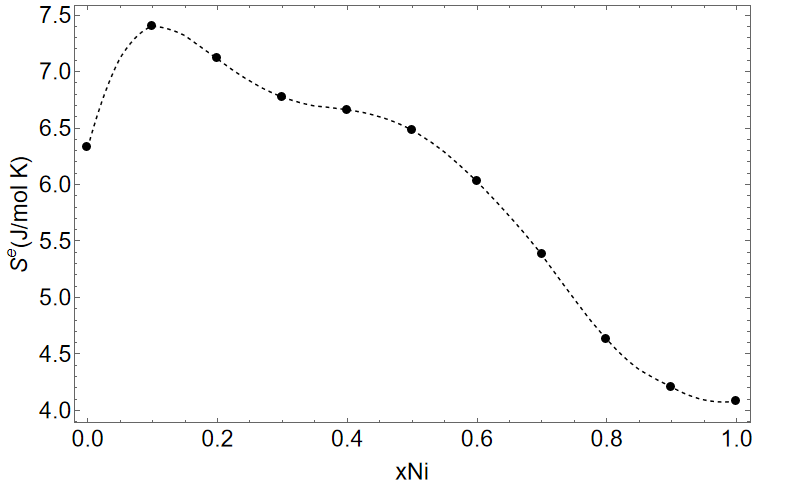}
    \caption{State entropy of Fe-Ni alloys evaluated at 1200 K from a multi-carrier model,  Lupis cluster data,  and the measured entropy data of Ono\cite{ono1977thermodynamic}.}
    \label{fig:nickel_entropy}
\end{figure}

The results are presented in Figure (\ref{fig:nickel_entropy}). The curve indicates that there is a large electronic contribution to the state entropy in Fe-Ni alloys as a function of temperature. This contribution is relatively flat as the Ni content increases until 65 at.$\%$ Ni. The cusp that was observed in the electronic entropy that coincided with the BCC/FCC transition in Figure (\ref{FeNielectronicentropyOneband}) disappears due to the continuous nature of the FCC solid solution at high-temperature. The trend in the high-temperature electronic entropy follows shares qualitative features with the compositional variation of the resistivity in Figure (\ref{FeNiIsothermalRes}). 
		
\subsubsection{Order-Disorder Transformation of FeNi$_{3}$}

\begin{figure}[h]			\includegraphics[width=\linewidth]{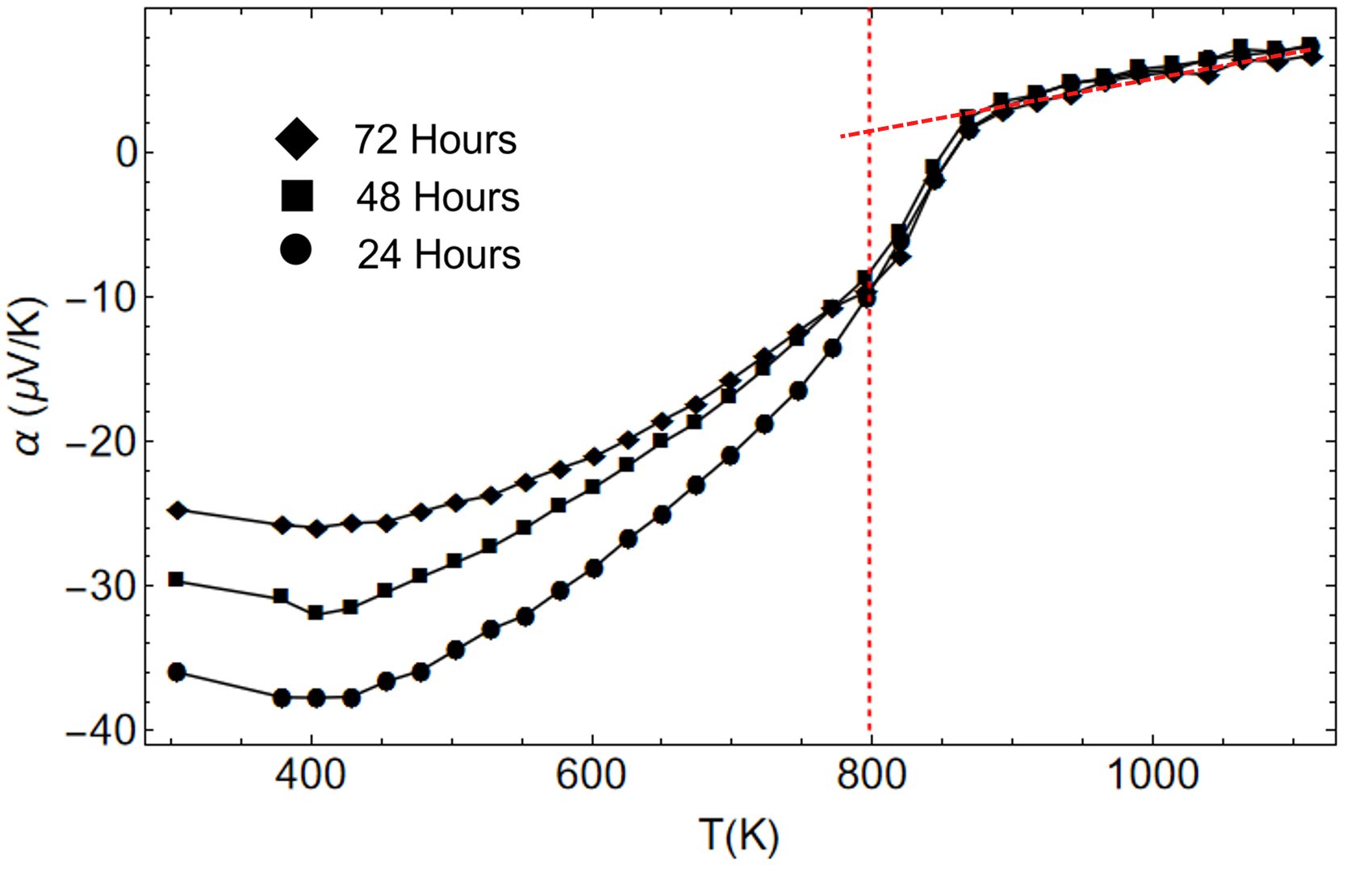}
				\caption{The absolute thermopower was measured as a function of temperature for samples annealed below the ordering temperature of FeNi$_{3}$ for up to 3 days. Increasing order is apparent in the thermopower below the ordering temperature.}
    \label{FeNi3ThermopowerData}
				\end{figure}

The temperature variation of the thermoelectric power measured for the FeNi$_{3}$ samples annealed below their ordering temperature (800K) for various durations prior to measurement is shown in Figure (\ref{FeNi3ThermopowerData}).  The thermopower converges at the transformation temperature,  regardless of the initial degree of ordering in the low-temperature phase.  We estimate the thermopower of the ordered phase to be 2 $\mu$V/K just below the phase-transition temperature thanks to the high-temperature slope to project back to the transformation temperature,  and estimate the thermopower of the fully disordered solid solution at 798 K to be -8.8 $\mu$V/K. We there assume the carrier concentration of Galepov and evaluate the electronic entropy change in a manner described in previous work \cite{paras2020electronic,paras2021contribution}. Our corresponding estimate of the electronic entropy for the order-disorder transition is -0.318 J/mol K.

Table (\ref{entropy_contributionsFeNi3}), compares this value with the total,  configurational and vibrational entropies values from others. The electronic entropy change is negative, which agrees with our previous work on the electron stabilization of ordered phases\cite{paras2021contribution}. But the contribution is much smaller in magnitude and therefore warrants discussion.

\begin{table*}
\begin{ruledtabular}
\begin{tabular}{cccccc}
Material  & T$^{c} $ (K) & $\delta S_{\text{config}}$ (J/molK) & $\delta S_{\text{vib}}$ (J/molK) & $\delta S_{\text{elec}}$ (J/molK) & $\delta S_{\text{T}}$ (J/molK) \\[0.2pt]
\hline 
\\[0.2pt]
FeNi$_{3}$ &  800  & 4.67582 & -1.55 & -0.318 & 3.44 \\[0.2pt]
\end{tabular}
\end{ruledtabular}
\caption{Various contributions to the entropy of the order-disorder transformation in FeNi$_3$ at a transition temperature near 800 K based on the total entropy of transition compiled by Hultgren [17].}
\label{entropy_contributionsFeNi3}
\end{table*}

\section{Discussion}

\subsection{Allotropic Transformation of Fe}

The electronic entropy of Fe presented in Table \ref{AllotropicTable} captures about 50\% of the entropy of the BCC to FCC transformation,  previously not accounted for.  This is rather surprising given the assumption made about the itinerant carrier concentration,  yet will be sustained qualitatively even with other assumptions about carriers.  The state electronic entropy will always be positive because the thermopower of $\gamma$FCC Fe is larger in magnitude than $\alpha$BCC (-6.8 vs 0.75 $\mu$V/K).  This agrees with the conjecture by Neuhaus et al that the electronic contribution is significant \cite{Neuhaus2014}. A more exact count of the free carrier concentration in Fe near the transition may be possible.  If we assume that the calculations presented for the vibrational entropy are correct, and that the electronic entropy accounts for the entire remaining fraction of the entropy,  a 1-band model would reproduce the appropriate electronic contribution to the allotropic entropy change. 

Prior work discussed interaction among electrons in d-bands\cite{ruderman1954indirect,Stearns1972} has been shown that not all d-band electrons are localized.  A fraction of these d electrons can be treated as if they were itinerant\cite{BethStearns1977,Stearns1972,Stearns1978,Stearns1973}.  Using the RKKY theory first proposed by Ruderman and Kittel,  an estimate of the number of such itinerant electrons in Fe, Ni and Co was found to be close to 0.4 d$_{i}$ per atom which is not too far from 0.6 d$_{i}$ per atom obtained here with a simplistic model. 

Modern computational efforts support the focus on the electron subsystem as being the main driver behind thermodynamic behavior.  Dynamical mean field theory (DMFT) methods have begun to quantitatively capture the phase transformation of transition metals as well as the formation of their magnetic properties,  where Density Functional Theory (DFT) and molecular dynamics (MD) have failed because they include the degrees of freedom of interacting electrons \cite{han2018phonon,lichtenstein2001finite}. That the entropy change should be at least 50$\%$ driven by the electron is therefore becoming an increasingly supported claim in other aspects of the field.   

Finally, these findings are fully substantiated by measurements of the Hall-effect through the BCC to FCC transition temperature.  Here, Okamato et al found that the ordinary Hall coefficient did not significantly change in magnitude after passing through an anomaly near the Curie temperature \cite{okamoto1962temperature}. Thus, unlike in the order-disorder case, in allotropic transitions, the distribution and partial entropies of the carriers may change but not the overall number of effective carries. This is also substantiated by the continuous nature of the resistivity measurements in the face of discontinuous changes in the thermopower.

\subsection{Fe-Ni Solid Solution Alloys}

		\subsubsection{300 K Solid Solution}
		Broad agreement between the electronic entropy derived from low-temperature heat capacity and the transport derived electronic entropy in the one-band assumption is observed  between 50 and 100 at.$\%$ Ni.  This is similar to the result that for the Cu-Ni alloy system,  and is rather unsurprising given the simple nature of the transport at room temperature in that region \cite{paras2024evidence}.
		
		Appeal to a two-band model is needed to account for the local maximum in the electronic entropy near in the INVAR, composition, Fe-30Ni.  There the thermopower tends toward zero and remains as such with temperature until the disordered high-temperature solid solution is formed (800K). 

		The formalism used herein provide constraints on to both the electron and hole distribution necessary to reproduce such calorimetric electronic entropies at room temperature.  This in an insight into the electronic structure of alloys at finite temperature that do not rely upon quantum oscillations.  The changes in concavity in the sub-band thermopowers, which are themselves measures of the entropy of each carrier type, substantiate the instability of the alloy near the INVAR composition. This is in agreement with both the change in sign in the enthalpy of mixing by Ono as well as the thermodynamic analysis by Tanji \cite{ono1977thermodynamic,tanji1978anomalous}.  They show that even upon the formation of a single FCC solution at high-temperature,  some small instability in the alloy may remain and may also account for the exotic properties of the INVAR composition. 

		\subsubsection{Central Atom Model of FCC solid-solution at 1200K}
		Discussing the inherent limitations of the model are beyond the scope of this work,  as the choice of this particular model is inherited form its deep anchoring into chemical bounds and ability to test the existence of local short-range order,  rather than its ability - or lack of - to reproduce phase boundaries.
		
		Yet it is important to note that the outcome of the calculated non-configurational entropy shown in Figure (\ref{entropycomparison}) is robust.  We found the outcome to be qualitatively rather independent on the choice of the cluster type and bond-energy or symmetry constraints that may be chosen during the construction of the cluster model. The results are also quite stable to fitting methods in this particular instance because the electronic state entropy of the end-members is large. The mixing contribution will become increasingly relatively important in systems where this is not true for one or both end-member chemistries. To be able to accurately minimize the number of fitting parameters and yet reproduce the changes in convexity of the experimental enthalpy of mixing  will require the ability to identify clusters at certain key stoechoimetric equivalent and control for the distribution of the experimental data that is being fit to.  Those clusters will not mix ideally and give rise to changes in convexity of the non-configurational entropy.  We find that experimentally  electronic state entropy measurements for such solution can support similar complex changes with composition, as discussed below.
		
		\subsubsection{1200K solid-solution}

	At high-temperature (1200 K) we cannot appeal to the calorimetric electronic entropy as a constraint on the electronic structure as proposed above for 300K.  The solution thermodynamic model required building clusters that reproduce the non-zero enthalpy of mixing and the entropy of mixing,  assuming an electronic entropy origin of the excess entropy. This was already demonstrated to be true in the Cu-Ni system, where despite the seemingly ideal phase diagram behavior, the entropy of mixing was built up from non-ideal configurational and electronic contributions\cite{paras2024evidence}. The result here was that the configurational entropy should be a much smaller fraction of the total measured thermodynamic entropy of mixing,  and this is observed again in Figure ({\ref{entropycomparison}). \cite{ono1977thermodynamic}. 

Whereas a second minimum in the magnitude of the thermopower emerges at 1200 K near 90 at.$\%$ Ni, this does not coincide with any other indication for a large electronic entropy such as a high resistivity, localization, or the emergence of peculiar phases like INVAR. Therefore we hypothesize that this is more indicative of changing the relative electron and hole mobility,  or their number,  as a function of temperature. 

The subband thermopowers that emerge in Figure (\ref{1200KSubbandSeebeck}) exhibit constant slopes near the INVAR composition,  not unlike the activity coefficients traversing a two-phase region.  Here however more than one carrier type is involved,  and that those types of carriers are distinguishable.  Observation of features that suggest solution mixing events among those carriers are reminiscent of earlier work that included the spatial mixing of electrons as a necessary to reproducing observed phase behavior in mixed solid oxides\cite{Zhou2006}.

Trends in electron and holes solution behavior at the phase level are noticeable.  In the case of Ni,  which has no allotropic transitions before melting, and a dominant electronic character,  the mixing behavior favors the FCC allotrope of iron for the solid solution.  Similar to the principle of "like-dissolves-like",  the electronic character of the end-members,  in the absence of dominant intermetallic compounds,  seems to determine the emergent phase behavior,  in a way that makes the crystal structure a consequence of the thermodynamics rather than the origin of the behavior itself. 

Outside of Cu-Ni, which has no such profound local maxima or minima in the thermopower far away from the equiatomic composition,  the maxima and minima in the resistivity and electronic entropy seem to coincide in Fe-Ni \cite{paras2024evidence,paras2023irreversible}. This implies that the multi-carrier interpretation we have proposed is necessary to describe the electronic entropy in that range of composition.  The only instance in which this may not be the case is if the increase in resistivity is driven by strong electron localization,  and that there is an electronic entropy associated with this that is captured by the calorimetric measurements but would not be captured by the present formalism.  Such configurational electronic entropies have been discussed in more detail elsewhere and would need to be revisited \cite{Zhou2006,VanVechten1973,Chandrashekhar1973}.

Future work should focus on both this distribution and the choice of band theory. The current framework is agnostic to how these are derived, and as the field advances, should serve as a useful beginning to more deeply understanding the electronic structure of metals at high-temperature. Hall-effect measurements will need to be instrumented at higher-temperatures to resolve some of the discrepancies in the proposed electron distribution and determine to what degree the system carrier concentration changes. Additional measurments will need to be undertaken as a function of atomic ordering.


\subsection{FeNi$_{3}$}

FeNi$_{3}$ electronic entropy decreases upon its disordering to a solid-solution. While the trend observed is similar to the findings in for Cu$_{3}$Au, the magnitude of the contribution is significantly smaller\cite{paras2021contribution}. In previous work on Cu-Au, it was found that the Hall-effect itself changed dramatically as a function of atomic ordering\cite{paras2021contribution}. In this instance, disordered Hall data were available,  which is suitable for the solid solution work, but it was found in Cu-Au that upon ordering, the magnitude of the Hall-effect decreased dramatically.  This raised the carrier concentration of the ordered phase resulting in a negative change in the electronic entropy upon disordering that was much larger in magnitude than observed here.  Therefore, this particular transition should be revisited when high-temperature Hall-effect measurements of the solid solution both near the disordering point and far enough below to exhibit fully ordered behavior,  are available.  For now, we may conclude that the electron subsystem also stabilizes the intermetallic phase of FeNi$_{3}$ relative to its solid solution.

\section{Conclusion}

This study has established a framework for using thermodynamic and electronic transport property measurements to understand the high-temperature electronic structure of metal alloys. This provides a new avenue to probe properties of the Fermi-surface that do not rely on quantum oscillation measurements or ab-initio methods.  Electronic entropy is found to play a significant role in the BCC to FCC transition in pure Fe, and shows the necessity of a multicarrier model to describe the electronic entropy in Fe-Ni alloys. The order-disorder transformation in FeNi$_{3}$ also highlights the importance of electronic contributions to entropy as a degree of ordering. This work demonstrates the need to instrument high-temperature Hall-effect measurements to fully realize the potential of using latent thermodynamic data to understand alloy electronic structures.

\section{Acknowledgements}
This material is based upon work supported by AFOSR under award number FA9550-20-1-0163. We would like to thank Professor David Clarke for the use of his ZEM-3.
\section{References}
\bibliography{main.bib}

{\clearpage}

\section{appendix}

\subsection{Central Atom Model}

We will first begin with the probabilities associated with different atomic configurations and thermodynamics functions.  We will use C.  Lupis notations as found in his textbook \cite{lupis1983chemical} throughout,  here for three types of clusters (central atom-based chemically possible configurations of A and B); A$_{3}$B, AB and AB$_{3}$, for a coordination number of 12,  as expected for the FCC lattice.  The choice of the composition of these culsters is inherited from the desire to minimize the number of fitting parameters - as explained below-,  while reproducing with maximum fidelity the experimental enthalpy of mixing results from Ono (Figure \ref{fig:CAMFit.}).  
In a random solution with a coordination number of Z,  the (*) probability of finding an A atom surrounded in the nearest neighbor shell by \(i\) atoms of B and \(Z-i\) atoms of A is 
\begin{equation}
  p^{(*\text{A})}_{i\text{B}} = C^Z_i X_\text{A}^{Z-i} X_\text{B}^i 
 \end{equation}
 
  \(C^Z_i\) is the combinatorial factor
   $$ \frac{Z!}{(Z-i)!i!} $$
   which is all the ways to randomly distribute such clusters by changing the choice of the first lattice site for the distribution.

The problem is that metallic solutions are not that random,  because they typically have enthalpies of mixing that are not zero.  This can be corrected by introducing \( f^\text{A}_{i\text{B}} \),  such that for a non-star probability (non-random solution): 
\begin{equation}
p^\text{A}_{i\text{B}} = p^{(*\text{A})}_{i\text{B}} \frac{f^\text{A}_{i\text{B}}}{P_\text{A}}
\end{equation}

where the probability of finding A in the solution is:
\begin{equation}
P_\text{A} = \sum_{i=0}^Z C^Z_i X_\text{A}^{Z-i} X_\text{B}^i f^\text{A}_{i\text{B}}
\end{equation}

and the actual probability of finding $i$ neighbors verifies the constraint:
\begin{equation}
\sum_{i=0}^Z p^\text{A}_{i\text{B}} = 1.
\end{equation}

The calculations by Lupis, ($J$ being either A or B),  lead to
\begin{equation}
\label{piBJ}
p_{i\text{B}}^J = \frac{C_Z^i X_\text{A}^{Z-i} X_\text{B}^i \exp \left( i\Lambda - \delta\psi_{i\text{B}}^J \right)}
{\sum_{i=0}^Z C_Z^i X_\text{A}^{Z-i} X_\text{B}^i \exp \left( i\Lambda - \delta\psi_{i\text{B}}^J \right)}
\end{equation}
where \(\Lambda\) is the Lagrange multiplier associated with a mass balance on all the atoms,  and 
\begin{equation}
\label{dpsiBJ}
\delta\psi_{i\text{B}}^J = \frac{\delta U_{i\text{B}}^J}{2RT} - \delta \ln q_{i\text{B}}^J
\end{equation}

is a bonding parameters that includes two contributions,  one from bonding ($\delta U_{i\text{B}}$) and one associated with the change in the internal partition function of the atom ($q_{i\text{B}}$) due to the new environment.

There are three equations to be satisfied:

\[
\Psi_1 = \sum_{i=0}^Z p_{i\text{B}}^\text{A} - 1 = 0
\]
\[
\Psi_2 = \sum_{i=0}^Z p_{i\text{B}}^\text{B} - 1 = 0
\]
\[
\Psi_3 = N_\text{A} \sum_{i=1}^Z i p_{i\text{B}}^\text{A} + N_\text{B} \sum_{i=1}^Z i p_{i\text{B}}^\text{B} - Z N_\text{B} = 0
\]

To solve further,  a functional form for the bonding parameter is proposed as a function of the cluster stoichiometry:
\begin{equation}
\delta \psi_{i\text{B}}^\text{A} = i \delta\psi_{1\text{B}}^\text{A} + \delta_{i,j'} \alpha' 
\end{equation}
\begin{equation}
\delta \psi_{i\text{B}}^\text{B} = i \delta\psi_{1\text{B}}^\text{B} + \delta_{i,j''} \alpha''
\end{equation}

where $\delta_{i,j'}$ and $\delta_{i,j'''}$ are the Kroneker symbols and $j'$ and $j''$ represent the tested configuration possible for a given cluster.  This is to describe that the energetic follows  linearly an increasing \(i\) (number of B nearest neighbors), until \(i = j\) where a perturbation by \(\alpha\) is found.  If \(\alpha\) is infinite and negative,  this causes a large negative \(\delta\psi\), increasing the probability of that particular cluster.  If \(\alpha\) is large and positive,  the particular cluster is energetically unfavored.  With our specific choice of 3 clusters, this becomes:

\begin{equation}
\begin{split}
\label{deltapsifunctions}
\delta\psi_{i \text{B}}^{\text{A}} = i \cdot  \delta\psi_{1\text{BA}} + \delta_{i,9} \alpha_1 + \delta_{i,6} \alpha_3 + \delta_{i,3} \alpha_5
\\
\delta\psi_{i \text{B}}^{\text{B}} = i \cdot \delta\psi_{1\text{BB}} + \delta_{i,9} \alpha_2 + \delta_{i,6} \alpha_4 + \delta_{i,3} \alpha_6
\end{split}
\end{equation}

We turn our attention now to the expression for the partition function, which is given by
\begin{equation}
Q = \overline{g} \, q_\text{A}^{N_\text{A}} q_\text{B}^{N_\text{B}} e^{-E/kT}
\end{equation}

and consequently,

\begin{equation}
G = -kT \ln Q \sim \overline{E} - kT \left( \ln \overline{g} + \ln \left( q_\text{A}^{N_\text{A}} q_\text{B}^{N_\text{B}} \right) \right).
\end{equation}

Statistically,  it is assumed that the partition function is highly peaked and that the average value of the degeneracy factor \(\overline{g}\) and the individual partition functions are dominating the sum to the point where they can be extracted from the integral as themselves.

Each atom A or B surrounded by other atoms has the potential energy \( U^\text{A}_{i\text{B}} \) or \( U^\text{B}_{i\text{B}} \).  The energy term may therefore be expressed as:
\begin{equation}
\begin{split}
E = \frac{1}{2} X_\text{A} \left( \sum_{i=0}^Z p^\text{A}_{i\text{B}} U^A_{i\text{B}} - U^A_{0\text{B}} \right) \\
+ \frac{1}{2} X_\text{B} \left( \sum_{i=0}^Z p^\text{B}_{i\text{B}} U^\text{B}_{i\text{B}} - U^\text{B}_{Z\text{B}} \right).
\end{split}
\end{equation}
where the probabilities are the ones that lead to the maximum term in the partition function as a sum.  Neglecting the \( PV \) term,   this leads to the enthalpy:

\begin{equation}
H^{\text{E}}_{\text{chem}} = E - X_\text{A} E_\text{A} - X_\text{B} E_\text{B}
\end{equation}

which is re-written in terms of the potential energies:

\begin{equation}
H^{\text{E}}_{\text{chem}} = E - \frac{1}{2} X_\text{A} U^\text{A}_{0\text{B}} - \frac{1}{2} X_ \text{B} U^\text{B}_{Z\text{B}}
\end{equation}

or

\begin{equation}
\begin{split}
H^{\text{E}}_{\text{chem}} = \frac{1}{2} X_\text{A} \left( \sum_{i=0}^Z p^\text{A}_{i \text{B}} U^\text{A}_{i\text{B}} - U^\text{A}_{0\text{B}} \right) \\
+ \frac{1}{2} X_\text{B} \left( \sum_{i=0}^Z p^\text{B}_{ i\text{B} } U^\text{B}_{i\text{B}} - U^\text{B}_{Z\text{B}} \right).
\end{split}
\end{equation}

The Gibbs energy is then:
\begin{equation}
G = H_{\text{chem}} - T S_{\text{conf}} - T S_{\text{nonconf}}
\end{equation}
where the internal partition function in the model contains all the non-configurational terms,  and the degeneracy factor contains all the atomic configurational terms according to:
\begin{equation}
\label{Sconf}
S_{\text{conf}} = k \ln g
\end{equation}

\begin{equation}
S_{\text{nonconf}} = k \left( \ln q_\text{A}^{N_\text{A}} + \ln q_\text{B}^{N_\text{B}} \right)
\end{equation}

Therefore,

\begin{equation}
S_{\text{nonconf}} = R \left( \sum_{i=0}^Z X_\text{A} p^\text{A}_{i\text{B}} \ln q^\text{A}_{i\text{B}} + \sum_{i=0}^Z X_\text{B} p^\text{B}_{i\text{B}} \ln q^\text{B}_{i\text{B}} \right)
\end{equation}

Similar to the findings with the excess enthalpy,  the excess entropy term is then equal to:

\begin{equation}
\begin{split}
S_{\text{nonconf}}^\text{E} = R \bigg( \sum_{i=0}^Z X_\text{A} p^\text{A}_{i\text{B}} \ln q^\text{A}_{i\text{B}} - X_\text{A} \ln q^\text{A}_{0\text{B}} \\
+  \sum_{i=0}^Z X_\text{B} p^\text{B}_{i\text{B}} \ln q^\text{B}_{i\text{B}} - X_\text{B} \ln q^\text{B}_{Z\text{B}} \Bigg)
\end{split}
\end{equation}

Defining:
\begin{equation}
\begin{split}
\delta \ln q^\text{A}_{i\text{B}} = \ln q^\text{A}_{i\text{B}} - \ln q^\text{A}_{0\text{B}} 
\\
    \delta \ln q^\text{B}_{i\text{B}} = \ln q^\text{B}_{i\text{B}} - \ln q^\text{B}_{Z\text{B}}
   \end{split}
\end{equation}
allows to re-write:

\begin{equation}
\begin{split}
\label{Snonconf}
S_{\text{nonconf}}^\text{E} = R \Bigg( X_\text{A} \sum_{i=0}^Z p^\text{A}_{i\text{B}} \delta \ln q^\text{A}_{i\text{B}} \\
+ X_\text{B} \sum_{i=0}^Z p^\text{B}_{i\text{B}} \delta \ln q^\text{B}_{i\text{B}}-X_\text{B} \delta \ln q^\text{B}_{Z\text{B}} \Bigg)
\end{split}
\end{equation}
%
%
The non-configurational entropy arises from the difference in the change in the energy of the internal partition function, 
and the change in the binding energy of the atoms,  as defined by $\delta \psi^J_{i\text{B}}$ in equation \ref{dpsiBJ}. 

Similarly defining:
\begin{equation}
\begin{split}
\delta U^\text{A}_{iB} = U^\text{A}_{i\text{B}} - U^\text{A}_{0\text{B}} 
\\
 \delta U^\text{B}_{i\text{B}} = U^\text{B}_{i\text{B}} -U^\text{B}_{Z\text{B}}
   \end{split}
\end{equation}

and setting $ \delta U^\text{B}_{Z\text{B}} $ at zero allows to recast the enthalpy according to:

 \begin{equation}
\begin{split}
\label{Hchemfit}
H_{\text{chem}}=E = \frac{1}{2} X_\text{A} \sum_{i=0}^Z p^\text{A}_{i \text{B}} \delta U^\text{A}_{i\text{B}} \\
+ \frac{1}{2} X_\text{B} \sum_{i=0}^Z p^\text{B}_{ i\text{B} } \delta U^\text{B}_{i\text{B}} - \frac{1}{2} X_\text{B} \delta U^\text{B}_{Z\text{B}}
\end{split}
\end{equation}

Herein,  using the 3 cluster configurations provides additional simplifications,  and the model
requires fitting only the probabilities and the values of $\delta\psi_{1\text{B}}^\text{A}$  and $\delta\psi_{1\text{B}}^\text{B}$.  The fitting is conducted in order to reproduce the enthalpy data trends with composition from Ono\cite{ono1977thermodynamic}. The probabilities are provided by equation \ref{dpsiBJ} with fitting of 6 parameters (the $\alpha_{i}$). The change in the binding energies is fitted as described in equation \ref{deltapsifunctions}.  The outcome of the fitting is presented in Table \ref{model_params}.

\begin{table}[h!]
\centering
\begin{tabular}{|c|c|}
\hline
\textbf{Parameter} & \textbf{Value} \\
\hline
$\alpha_1$ & 0.3 \\
$\alpha_2$ & 0.3 \\
$\alpha_3$ & -0.9 \\
$\alpha_4$ & -0.9 \\
$\alpha_5$ & -0.6 \\
$\alpha_6$ & -0.6 \\
$\delta\psi_{1\text{B}}^\text{A}$ & -0.15 \\
$\delta\psi_{1\text{B}}^\text{B}$ & -0.2 \\
\hline
\end{tabular}
\caption{Central Atom Model Parameters}
\label{model_params}
\end{table}

\FloatBarrier 

\clearpage

\onecolumngrid

\subsection{X-Ray Diffraction of Solid Solution Fe-Ni Alloys}
	\begin{figure*}[h]
	\includegraphics[width=0.8\linewidth,keepaspectratio]{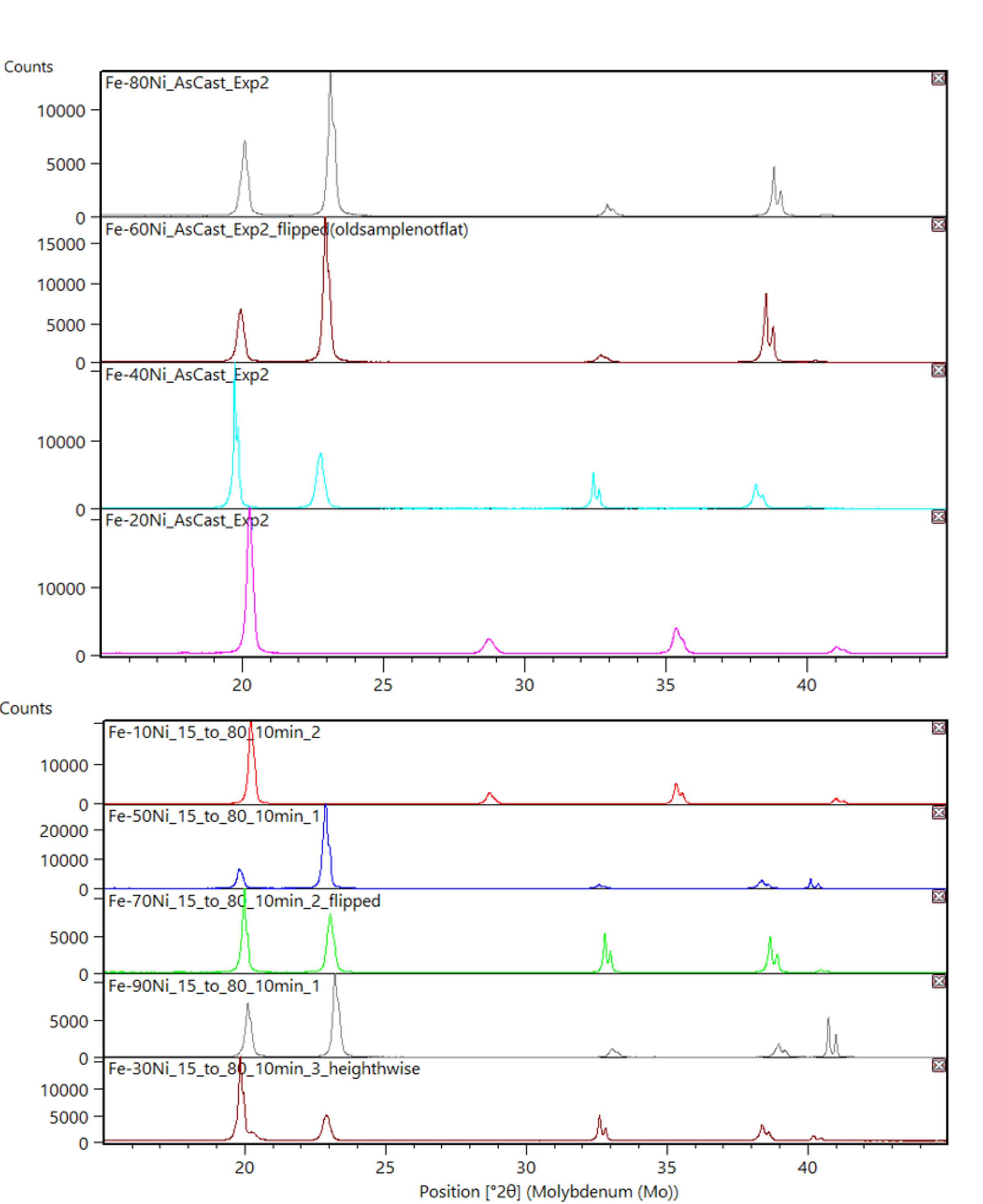}
				\caption{The XRD patterns were measured for Fe-Ni alloys to ensure their homogeneity. These measurements were conducted in an Panalytical Empyrean with Mo-K$\alpha$ radiation in a 2$\theta$, Bragg-Brentano configuration. Additional experiments were conducted in a Rigaku smart-lab using Cu-K$\alpha$ radiation to ensure that the observed peak splitting was not due to a secondary phase.}
					\label{XRDFeNi}
				\end{figure*}

\clearpage

    \subsection{Proposed Electron and Hole Distributions at 300 K and 1200 K in Fe-Ni Alloys}

    \begin{figure*}[h]
				\includegraphics[width=0.70\linewidth,keepaspectratio]{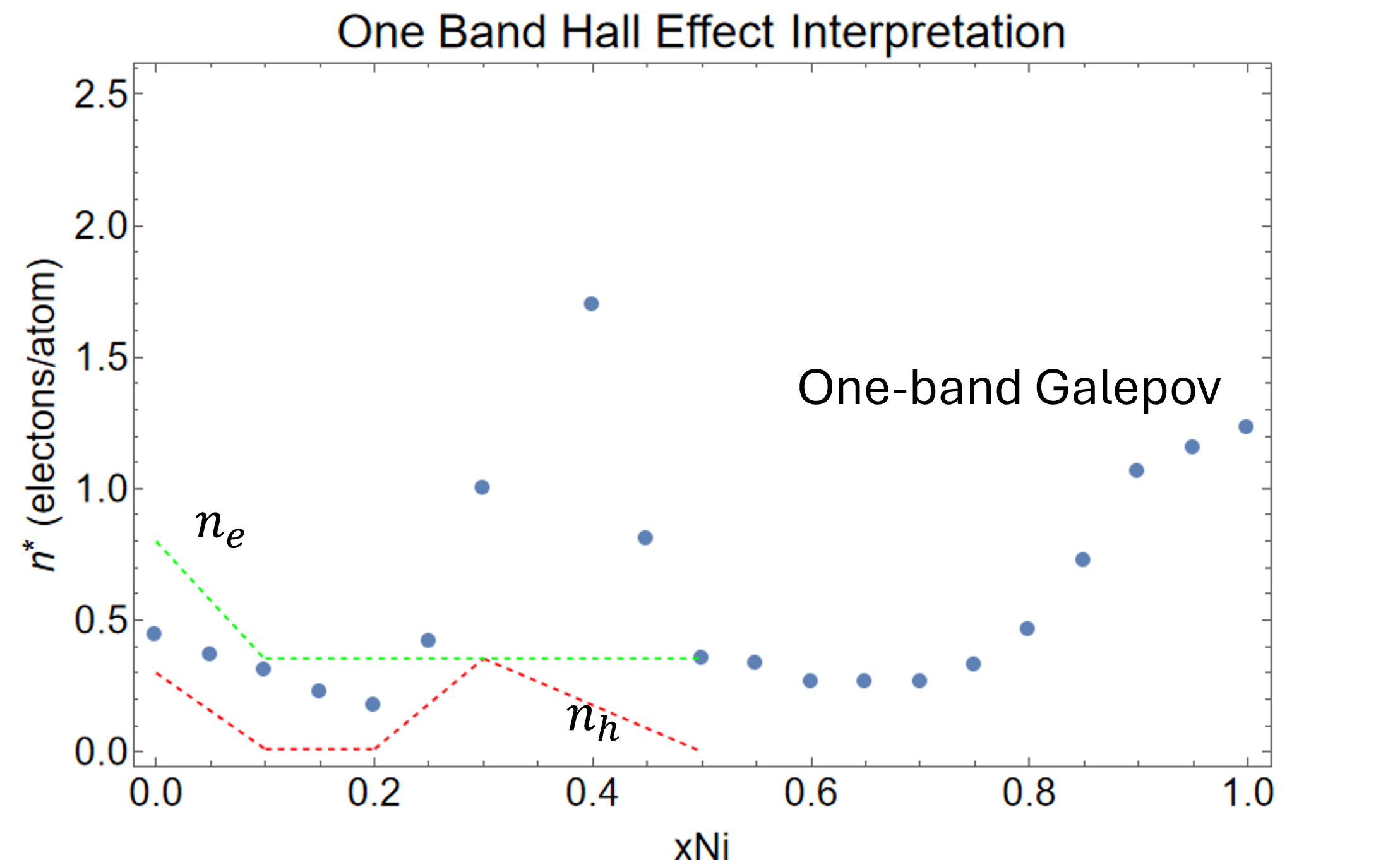}
				\caption{Proposed electron and hole distribution to fit a two-band model that agrees with thermodynamically derived calorimetric electronic entropies at 300K in Fe-Ni alloys. The values of Stearns were used in the two-band model for Fe, electron and hole number were equal at 30 at.$\%$ Ni, and then the number of holes was sent to zero as the Ni concentration approached 50 at. $\%$ Ni and obtained the one-band assumption value derived from Hall-effect data owing to the success of the single band model in reproducing the calorimetric electronic entropy\cite{stearns1971origin,stearns1973origin}.}
    				\label{300K Electron Distribution}
				\end{figure*}

\begin{figure*}[h]
				\includegraphics[width=0.70\linewidth,keepaspectratio]{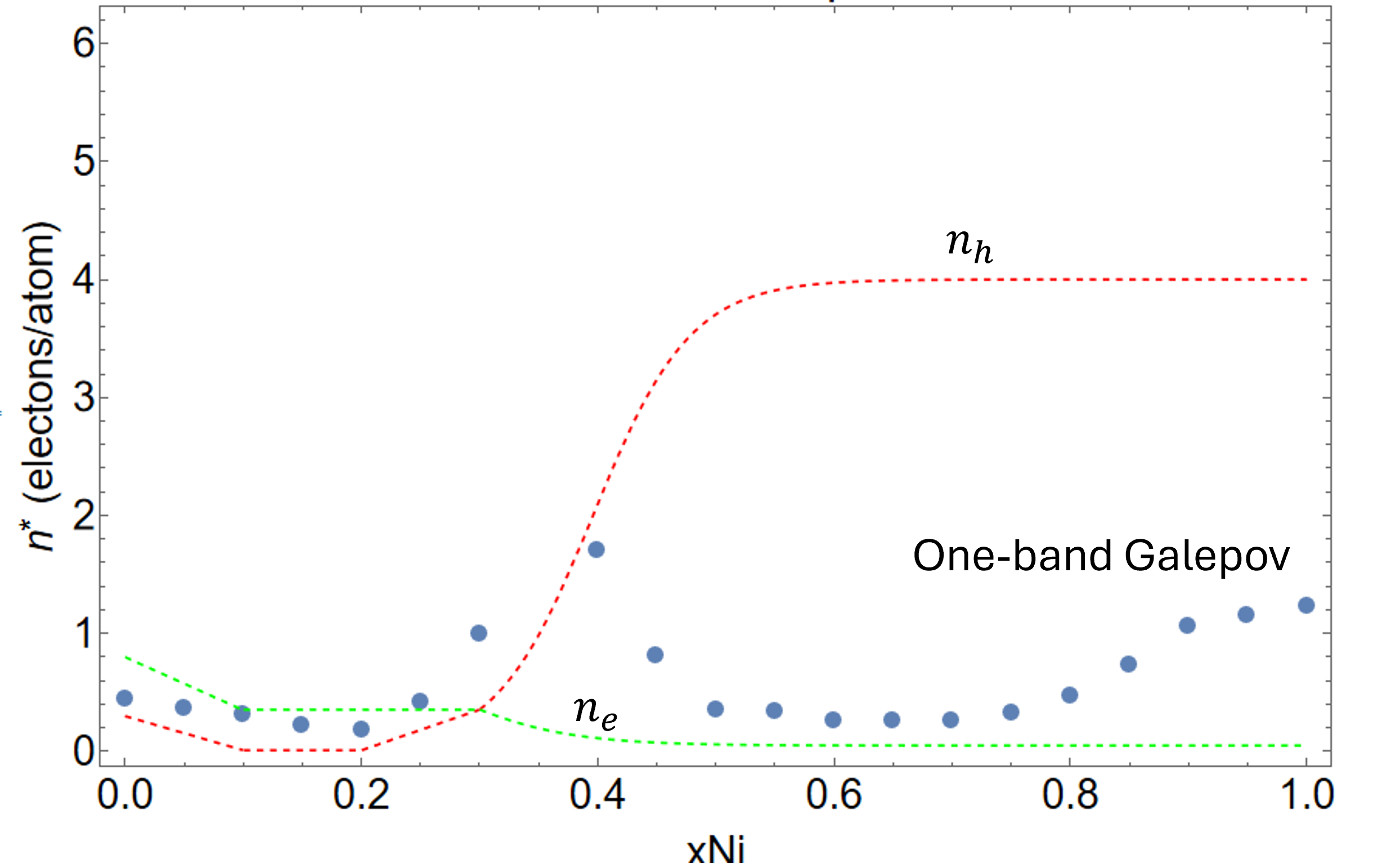}
				\caption{Proposed electron and hole distribution to evaluate two-band model that agrees with thermodynamically derived excess electronic entropies at 1200K in Fe-Ni alloys. Identical assumptions up to 30 at.$\%$ Ni at 300K were made, however it was necessary to increase the hole concentration considerably using a logistic function because of the change in sign of the thermopower at high-temperature in Fe-Ni alloys in order to obtain self-consistent electron and hole thermopowers.}
    				\label{1200K Electron Distribution}
				\end{figure*}

\subsection{Subband Thermopowers}

\clearpage
\begin{figure*}
    \centering
    \includegraphics[width=\textwidth]{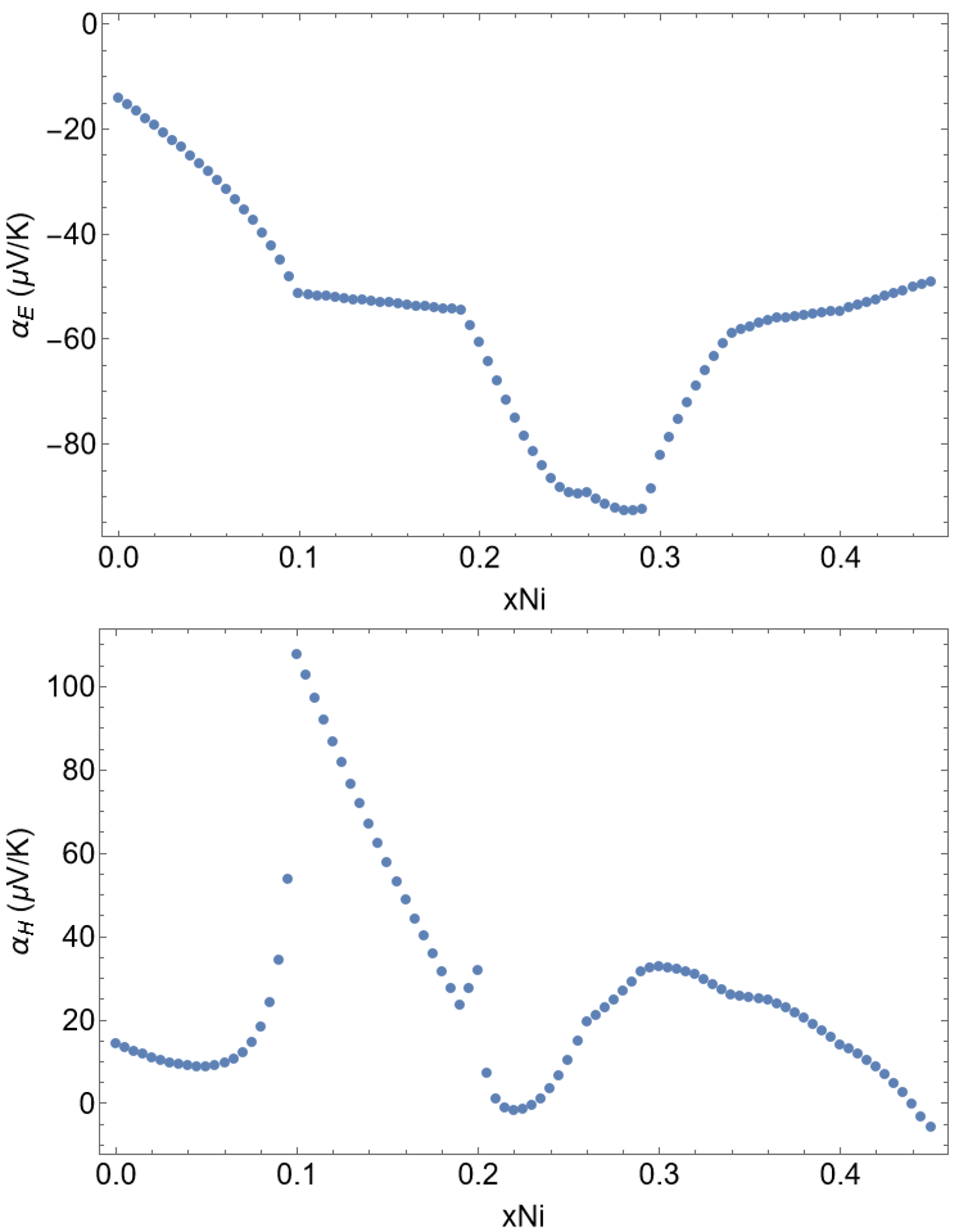}
    \caption{The electron ($\alpha_{E}$) and hole ($\alpha_{H}$) subband thermopower coefficients proposed to reproduce the electronic entropy of Fe-Ni at 300 K. }
\label{RoomTemperatureSubbandSeebeck}
\end{figure*}

\clearpage
\begin{figure*}
    \centering
\includegraphics[width=\textwidth]{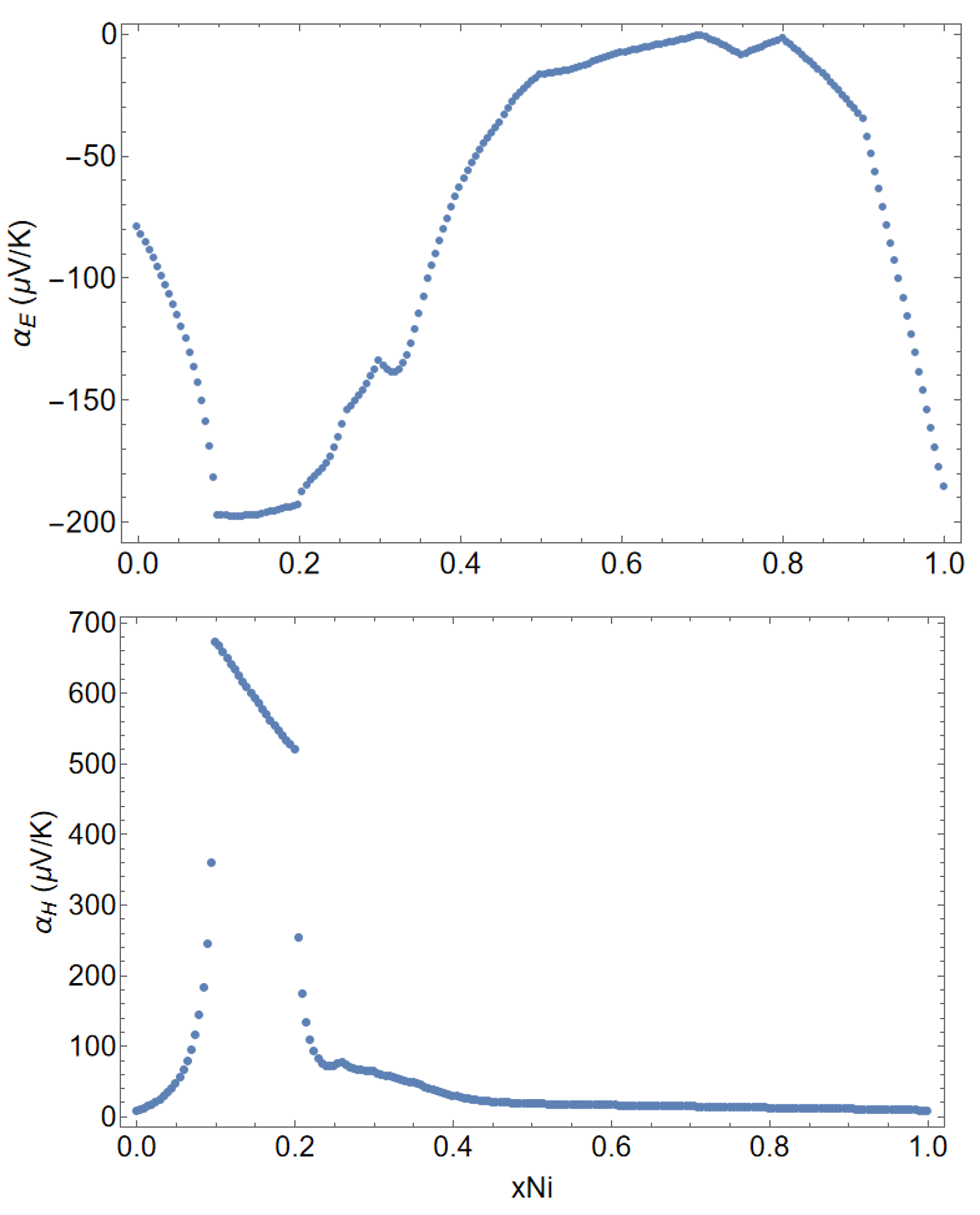}
    \caption{The electron ($\alpha_{E}$) and hole ($\alpha_{H}$) subband thermopower coefficients proposed to reproduce the electronic entropy of Fe-Ni at 1200 K. }
    \label{1200KSubbandSeebeck}
\end{figure*}
\clearpage

\subsection{One-band electronic entropy}

   \begin{figure*}[h]
    \centering
\includegraphics[width=\textwidth]{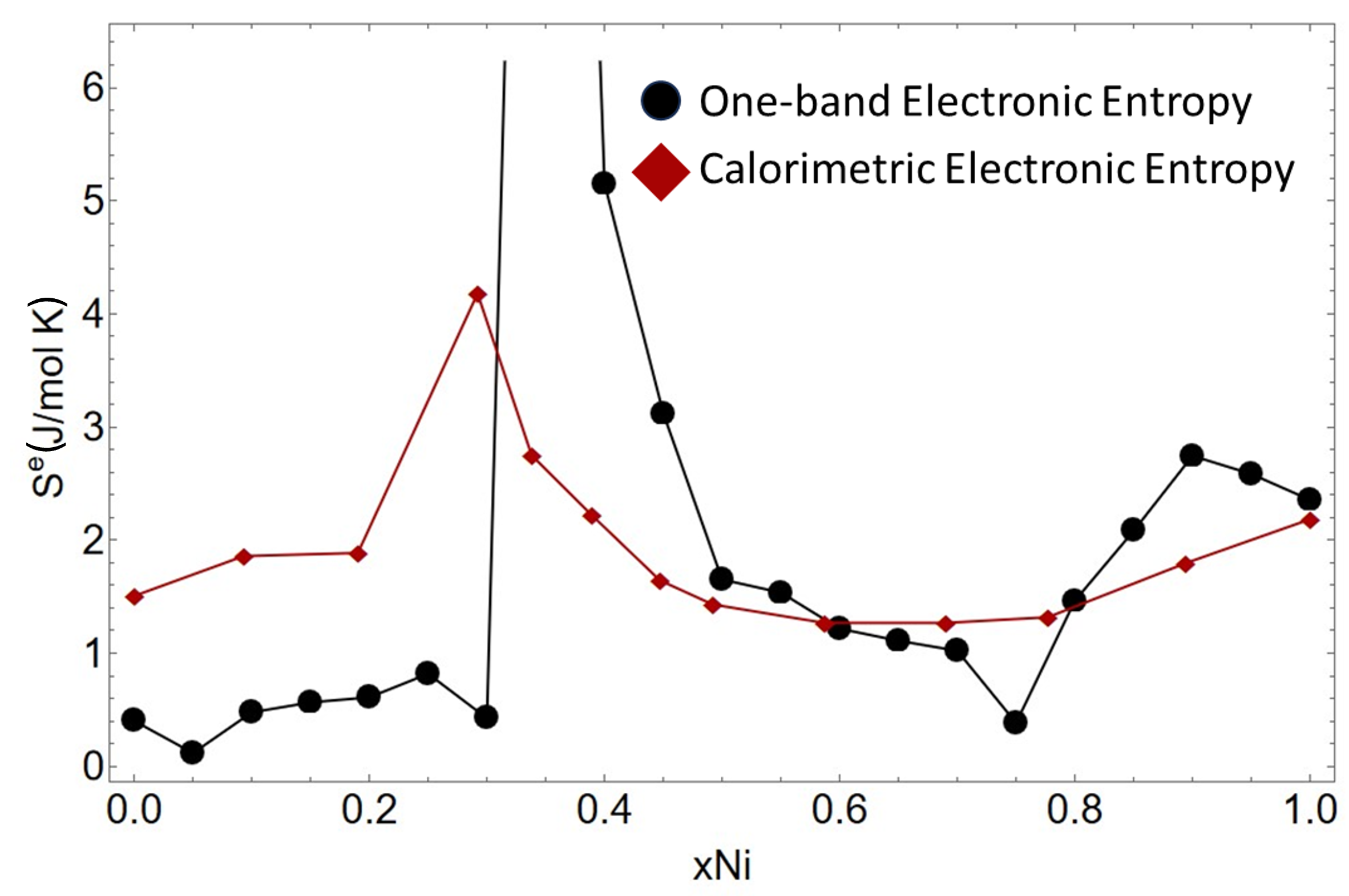}
    \caption{This is a figure of the electronic entropy derived from applying a simple one band assumption to the Hall-effect, compared to the electronic heat capacity. Note the broad agreement above 50 at.$\%$ Ni, below which a two-band model is used to rationalize the electronic entropy to the one provided by the electronic heat capacity.}
    \label{OneBandCalorimetricComparison}
\end{figure*} 
\end{document}